\documentclass[10pt,journal,compsoc]{IEEEtran}
\ifCLASSOPTIONcompsoc
\usepackage[nocompress]{cite}
\else
\usepackage{cite}
\fi

\ifCLASSINFOpdf
\else
\fi

\usepackage{braket}
\usepackage{xspace}
\usepackage{stfloats}
\usepackage{amsmath,amsfonts}
\usepackage{graphicx}
\usepackage{textcomp}

\def\BibTeX{{\rm B\kern-.05em{\sc i\kern-.025em b}\kern-.08em
T\kern-.1667em\lower.7ex\hbox{E}\kern-.125emX}}

\usepackage{hyperref}
\hypersetup{
colorlinks = true, 
linkcolor=black, 
citecolor=black, 
urlcolor=black 
}

\usepackage{paralist}
\usepackage{subcaption}
\usepackage{pifont}
\usepackage[most,breakable]{tcolorbox}
\usepackage{booktabs}
\usepackage{xspace}
\usepackage{bbding}
\usepackage{multirow}
\usepackage{multicol}
\usepackage{amsthm}
\usepackage{caption}
\usepackage{xcolor,colortbl}
\usepackage{subcaption}
\usepackage{tabularx}
\usepackage{siunitx}

\theoremstyle{definition}
\newtheorem{definition}{\protect\definitionname}
\providecommand{\definitionname}{Definition}
\theoremstyle{definition}
\newtheorem{example}{\protect\examplename}
\providecommand{\examplename}{Example}
\theoremstyle{definition}
\newtheorem{remark}{Remark}

\newcommand{\totalVariable}{\ensuremath{N}\xspace}
\newcommand{\binaryVariables}{\Vec{z}\xspace}
\newcommand{\variable}[1]{z_{#1}\xspace}
\newcommand{\spin}[2]{\sigma_{#1}^{#2}\xspace}
\newcommand{\IsingObj}[1]{C(#1)\xspace}
\newcommand{\proHam}{H_C\xspace}
\newcommand{\mixHam}{H_M\xspace}
\newcommand{\proUni}[1]{U_C(#1)\xspace}
\newcommand{\mixUni}[1]{U_M(#1)\xspace}
\newcommand{\expEq}[2]{E_C(#1,#2)}
\newcommand{\ourApproach}{\textcolor{black}{IGDec-QAOA}\xspace}
\newcommand{\baselineqaoa}{\textsc{Div-QAOA}\xspace}

\newcommand{\method}{\textit{BootQA}\xspace}

\newcommand{\testCase}[1]{\ensuremath{t_{#1}}\xspace}
\newcommand{\totalTestCase}{\ensuremath{n}\xspace}
\newcommand{\attrNum}{\ensuremath{m}\xspace}

\newcommand{\effAttrValueSet}[1]{\ensuremath{\mathbf{v_{#1}}}\xspace}
\newcommand{\costAttrValueSet}[1]{\ensuremath{\mathbf{c_{#1}}}\xspace}
\newcommand{\effValue}[2]{\ensuremath{v_{#1}^{#2}}\xspace}
\newcommand{\costValue}[2]{\ensuremath{c_{#1}^{#2}}\xspace}

\newcommand{\weight}[1]{\ensuremath{w}_{#1}\xspace}
\newcommand{\testCaseSet}{\ensuremath{T}\xspace}
\newcommand{\seltestCaseSet}{\ensuremath{T'}\xspace}

\newcommand{\pa}{\textit{Paint Control}\xspace}

\newcommand{\objFunc}{\ensuremath{\mathcal{O}}}

\newcommand{\obj}[1]{\ensuremath{f_{#1}}}
\newcommand{\objSet}{\ensuremath{F}}

\newcommand{\subproblemSize}{\ensuremath{N}\xspace}

\newcommand{\numEffAttr}{\ensuremath{s}\xspace}
\newcommand{\numCostAttr}{\ensuremath{q}\xspace}
\newcommand{\activeSet}{\textit{actSet}\xspace}
\newcommand{\fractionNum}{\ensuremath{\mathit{num}}\xspace}
\newcommand{\appro}{\ensuremath{\mathit{ar}}\xspace}
\newcommand{\extime}{\textit{exTime}\xspace}

\newcommand{\numEval}{\textit{numEva}\xspace}

\newcommand{\pop}{\textit{popSize}\xspace}
\newcommand{\const}{\ensuremath{\mathit{Const}}\xspace}
\newcommand{\Atwelve}{\ensuremath{\hat{A}}\textsubscript{12}\xspace}
\newcommand{\etal}{\textit{et al.}\xspace}

\newcommand{\revision}[1]{\textcolor{black}{#1}}

\begin{document}
%

\title{\revision{Quantum Approximate Optimization Algorithm for Test Case Optimization}}


%
%
%
%

\author{Xinyi Wang, Shaukat Ali, Tao Yue, and Paolo Arcaini
\IEEEcompsocitemizethanks{\IEEEcompsocthanksitem Xinyi Wang is with Simula Research Laboratory, Oslo, Norway and University of Oslo, Oslo, Norway.
\IEEEcompsocthanksitem Shaukat Ali is with Simula Research Laboratory, Oslo, Norway and Oslo Metropolitan University, Oslo, Norway.
\IEEEcompsocthanksitem Tao Yue is with Simula Research Laboratory, Oslo, Norway.
\IEEEcompsocthanksitem Paolo Arcaini is with National Institute of Informatics, Tokyo, Japan.}
\thanks{Manuscript received April 19, 2005; revised August 26, 2015.}}

\markboth{IEEE Transactions on Software Engineering}%
{Wang \MakeLowercase{\textit{et al.}}: Quantum Approximate Optimization Algorithm for Test Case Optimization}

%



\IEEEtitleabstractindextext{%
\begin{abstract}
\revision{Test case optimization (TCO) reduces the software testing cost while preserving its effectiveness. However, to solve TCO problems for large-scale and complex software systems, substantial computational resources are required. Quantum approximate optimization algorithms (QAOAs) are promising combinatorial optimization algorithms that rely on quantum computational resources, with the potential to offer increased efficiency compared to classical approaches. Several proof-of-concept applications of QAOAs for solving combinatorial problems, such as portfolio optimization, energy optimization in power systems, and job scheduling, have been proposed. Given the lack of investigation into QAOA's application for TCO problems, and motivated by the computational challenges of TCO problems and the potential of QAOAs,}
we present \ourApproach to formulate a TCO problem as a QAOA problem and solve it on \revision{both ideal and noisy quantum computer simulators, as well as on a real quantum computer.} To solve bigger TCO problems that require many qubits, which are unavailable these days, we integrate a problem decomposition strategy with the QAOA. We performed an empirical evaluation with five TCO problems and four publicly available industrial datasets from ABB, Google, and Orona to compare various configurations of \ourApproach, assess its decomposition strategy of handling large datasets, and compare its performance with classical algorithms (i.e., Genetic Algorithm (GA) and Random Search). Based on the evaluation results achieved on an ideal simulator, we recommend the best configuration of our approach for TCO problems. Also, we demonstrate that our approach can reach the same effectiveness as GA and outperform GA in two out of five test case optimization problems we conducted. \revision{In addition, we observe that, on the noisy simulator, \ourApproach achieved similar performance to that from the ideal simulator. Finally, we also demonstrate the feasibility of \ourApproach on a real quantum computer in the presence of noise.}
\end{abstract}

}

\maketitle

\IEEEdisplaynontitleabstractindextext

%
\IEEEpeerreviewmaketitle

\IEEEraisesectionheading{\section{Introduction}\label{sec:introduction}}
\revision{\IEEEPARstart{T}{o} ensure the quality and reliability of software systems, software testing is extremely vital in the software development life cycle. However, with the growing size and complexity of software systems, traditional testing approaches, such as unit testing and integration testing, face challenges regarding effectiveness and efficiency since a large number of test cases need to be executed. Indeed, such test case execution is always limited by available time and resource budgets. This leads to the development of test case optimization (TCO) techniques, which aim to reduce the number of test cases to execute to minimize the testing cost (e.g., execution time) while preserving its effectiveness (e.g., code coverage). TCO problems are combinatorial optimization problems and are NP-hard~\cite{bajaj2019systematic, pan2023atm}. Thus, with the increasing software scale, even larger classical computational resources cannot explore the significant part of the vast search space. Consequently, traditional TCO approaches face computational challenges when optimizing test cases for classical software systems.}

\revision{Quantum Computing (QC) is a novel computational paradigm that can potentially tackle these computational challenges. QC has been expanding in recent years, with the potential to perform certain complex calculations faster than classical computers by leveraging quantum mechanics principles. QC's applications in various domains (e.g., finance, complex physics and chemistry simulations, and drug discovery~\cite{motta2022emerging, zinner2021quantum, egger2020quantum}) have emerged. Among all these potential applications, optimization applications are expected to demonstrate a quantum advantage in the near future, especially for solving combinatorial optimization problems. Quantum Approximate Optimization Algorithm (QAOA) is one of the promising algorithms~\cite{farhi2014quantum,QAOA} for solving such problems. It is a hybrid algorithm that contains a variational quantum circuit and a classical optimizer, to explore large search space efficiently, and approximate the optimal solution of a combinatorial optimization problem. QAOA applications have increasingly emerged and demonstrated the capability of solving optimization problems, such as scheduling~\cite{QAOAScheduling}, portfolio optimization~\cite{QAOAPorfolio}, and optimization in power systems~\cite{QAOAPower}.}

\revision{To deal with the computational challenges of TCO problems and demonstrate the potential of QAOA as a possible solution, we present the first application of QAOAs for solving TCO problems, i.e., test case selection (TCS) and test case minimization (TCM) problems in the context of testing classical software systems. In particular, we propose an approach named \ourApproach, based on QAOAs, to solve TCO problems, by making the following contributions.} First, we demonstrate how to formulate optimization objectives for TCO problems as a generic function in the Ising formulation~\cite{lucas2014ising} such that QAOAs can be applied to solve them. Second, to solve problems with a large number of test cases that cannot be directly solved with current quantum computers having a limited number of qubits, we apply an impact-guided decomposition strategy adapted from D-Wave~\cite{dwavePartitioning} to break down a complex TCO problem into a set of smaller and more manageable sub-problems so that they can be solved with current quantum computers.

Third, we perform an empirical evaluation with five publicly available industrial case studies from ABB, Google, and Orona to demonstrate \ourApproach's capability of finding approximate optimal solutions for TCO problems. In particular, we study:
\begin{inparaenum}[(a)]
\item Which configurations of \ourApproach are the best for solving TCO problems?
\item Does the impact-guided decomposition of \ourApproach contribute to its performance?
\item How is the performance of \ourApproach as compared to classical algorithms, i.e., Genetic Algorithm (GA) and Random Search (RS)? 
\revision{\item How is the performance of \ourApproach on a noisy quantum computer?}
\end{inparaenum}
\revision{Given that current quantum computers inherently suffer from noise in their computations, we also compare the results of \ourApproach on the ideal simulator with that of the noisy simulator to evaluate the impact of noise. Finally, we assess the feasibility of \ourApproach on IBM's quantum computer by solving a small-scale TCM problem in the presence of noise.}

Based on the results of our empirical evaluation on the ideal simulator, we select the best configurations of \ourApproach to solve the TCO problems. Our results show that the impact-guided decomposition strategy of \ourApproach indeed contributes to its performance in finding optimal solutions. In addition, the effectiveness of our approach is on par with GA, with 2 (out of 5) TCO problems outperforming GA. This demonstrates the benefit of using \ourApproach for solving TCO problems with currently available small-scale quantum computers. \revision{In addition, the results of the experiment comparing \ourApproach on ideal and noisy simulator showed that the performance of \ourApproach on noisy simulator is comparable to that on the ideal simulator. Regarding the experiment on IBM's quantum computer for solving a small-scale TCM problem, \ourApproach also reached the optimal solution obtained by the ideal simulator, thereby providing evidence about the feasibility of running \ourApproach on real quantum computers. These findings also demonstrate that the impact of noise on \ourApproach is insignificant.} We envision a further improvement of \ourApproach's performance in the future when large-scale quantum computers will be available. We provide detailed experiment results, code, and data in the repository: \url{https://doi.org/10.5281/zenodo.13911651}.

\begin{remark}
This paper does not intend to demonstrate the quantum advantage of solving TCO problems with QAOA over classical approaches. This is practically impossible, since only small-scale noisy quantum computers are currently available. Rather, we demonstrate how TCO problems can be formulated and solved with QAOA by achieving comparable performance as classical approaches with manageable execution costs, which should be considered as a cornerstone for future quantum computing applications in solving software engineering optimization problems.
\end{remark}

\section{Background} \label{sec:background}
\subsection{\revision{Test Case Optimization}} \label{subsec:tco}
\revision{Delivering software on time while ensuring its reliability is a major challenge in the software industry. It is necessary to conduct software testing efficiently while keeping high effectiveness; to this aim, test case selection (TCS) and test case minimization (TCM) are commonly used techniques that choose a test suite from a test case pool of software under test (SUT). The main goal of TCS is to ensure that the most relevant and important tests are run, optimizing resource use while maintaining test effectiveness. TCM focuses on eliminating redundant or non-essential test cases while maintaining essential functionality and coverage. Both techniques aim to solve combinatorial optimization problems, seeking to find tradeoff solutions among several conflicting objectives, such as execution time, code coverage, and failure rate.}

\subsection{Quantum Computing} \label{subsec:qc}
Quantum Computing (QC) utilizes quantum bits (\textit{qubits}) for performing computations. Qubits are different from classical bits since a classical bit can be in a state of either 0 or 1 at a time, \revision{while each qubit can exist in a superposition of both states until measurement, which then collapses the qubit to a definite state of either 0 or 1, resulting in a classical bit.} A one-qubit quantum state can be represented in the Dirac notion as: $\ket{\psi} = \alpha_0\ket{0}+\alpha_1\ket{1}$, where $\alpha_0$ and $\alpha_1$ represents the \textit{amplitude} of the quantum state and is a complex number. $|\alpha_0|^2$ and $|\alpha_1|^2$ determine the probabilities of being in states $\ket{0}$ and $\ket{1}$ respectively when measured. The sum of $|\alpha_0|^2$ and $|\alpha_1|^2$ shall be equal to 1. Analogously to classical logic gates used in classical computers, we use quantum gates operated on qubits to process quantum information. These quantum gates are represented by unitary matrices, also called unitary operators. We provide quantum gate descriptions in Table~\ref{table:gatetype} that are relevant to building QAOA quantum circuits.
\begin{table}[!tb]
\small
\caption{Descriptions of gates used in QAOA}
\resizebox{\columnwidth}{!}{
\begin{tabular}
{m{0.25\columnwidth}|m{0.7\columnwidth}}
\toprule
\textbf{Gate} & \textbf{Description} \\
\midrule
\textit{Hadamard (H)} & It puts a qubit into an equal superposition, i.e., giving an equal chance (i.e., 50\%) of being in state $\ket{0}$ or $\ket{1}$, when measured. Applying an H gate on a qubit with $\ket{0}$ state creates a \revision{state of $\frac{1}{\sqrt{2}}\ket{0}+\frac{1}{\sqrt{2}}\ket{1}$, which can be denoted as} $\ket{+}$ state.\\ \hline
\textit{Pauli-z} & It rotates a qubit around the z-axis with $\pi$ radians represented as $\spin{}{z}$. \\ \hline
\textit{Pauli-x} & It rotates a qubit around the x-axis with $\pi$ radians. It can be represented as $\spin{}{x}$. \\ \hline
\textit{Rz($\theta$)} & It rotates a qubit around the z-axis with $\theta$ radians. It is represented as $e^{-i\theta/2\spin{}{z}}$. \\ \hline
\textit{Rx($\theta$)} & It rotates a qubit around the x-axis with $\theta$ radians. It can be represent as $e^{-i\theta/2\spin{}{x}}$. \\ \hline
\textit{Controlled-NOT (C-NOT)} & It is a two-qubit gate consisting of a control qubit and a target qubit. If the control qubit is in state $\ket{1}$, the target qubit rotates around the x-axis with $\pi$ radians. \\ \bottomrule
\end{tabular}
}
\label{table:gatetype}
\end{table}
%

\subsection{Quantum Approximate Optimization Algorithm (QAOA)}\label{subsec:qaoabackground}
\begin{figure}[!tb]
\centering
\includegraphics[width=\columnwidth]{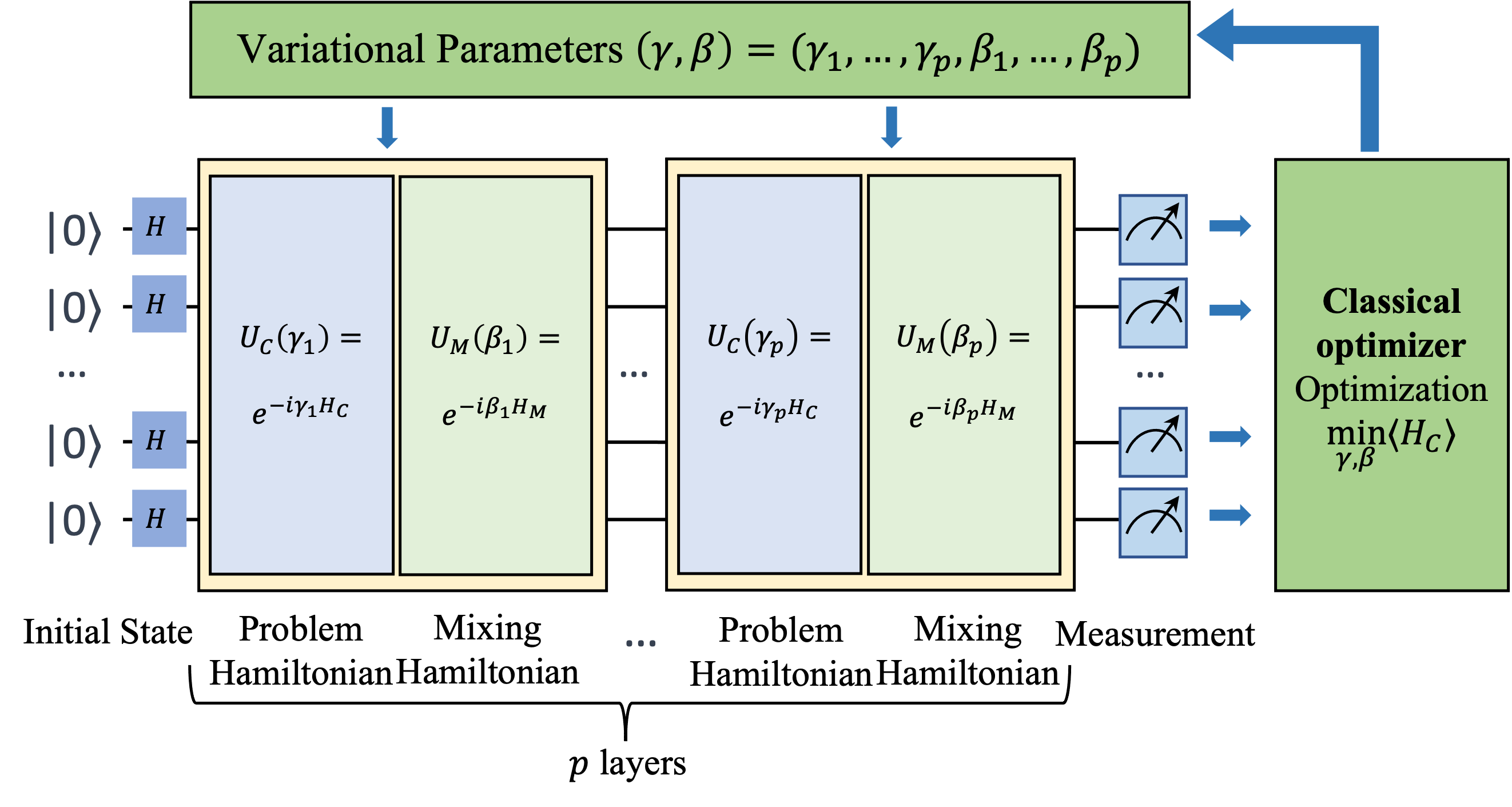}
\caption{A typical configuration of QAOA \revision{(\textit{p} is the depth)}}
\label{fig:qaoa}
\end{figure}

QAOA is a promising hybrid quantum-classical algorithm proposed to tackle combinatorial optimization problems\revision{~\cite{farhi2014quantum, QAOA}}. It is a variational quantum algorithm that aims to find the approximate optimal solution to an optimization problem. A parameterized quantum circuit is at the core of QAOA, whose parameters are optimized with a classical optimizer. The objective of the parameterized circuit is to approximate the adiabatic evolution of a quantum system, starting with an initial Hamiltonian\footnote{Hamiltonian is a Hermitian operator that describes the quantum state evolution in a quantum system~\cite{nielsen2002quantum}.} and reaching the ground state that represents the solution, i.e., the final Hamiltonian. Fig.~\ref{fig:qaoa} shows a typical configuration of QAOA.

\revision{To encode a combinatorial optimization problem for QAOA, an objective function should be defined in the form of an Ising formulation~\cite{lucas2014ising}. Such formulation is expressed as a quadratic formulation with \totalVariable binary variables, denoted by $\variable{i}$, which takes values of either +1 or -1.}
%
\revision{
\begin{equation} \label{eq:IsingObj}
\IsingObj{\binaryVariables}=\sum_{i=1}^{\totalVariable}h_i\variable{i}+\sum_{i=1}^{\totalVariable}\sum_{j=i+1}^{\totalVariable}J_{i, j}\variable{i}\variable{j},\quad \variable{i} \in \{+1,-1\}
\end{equation}
}

In Eq.~\ref{eq:IsingObj}, $\IsingObj{\binaryVariables}$ is the objective function, and $\binaryVariables$ represents the bit string. \revision{$h_i$ is the linear coefficient and $J_{i, j}$ represents the quadratic coefficient.} With this objective function, a classical problem can be converted into a problem Hamiltonian that can be solved by QAOA.

As depicted in Fig.~\ref{fig:qaoa}, QAOA has the following three main components.

\noindent\textbf{Initial state.}
It is the starting point for the evolution of a quantum system represented by a quantum circuit, where each qubit represents a variable in Eq.~\ref{eq:IsingObj}. 
Commonly, we initialize the quantum system into an equal superposition of all its qubits by applying Hadamard gates (detailed description is shown in Table~\ref{table:gatetype}) on all qubits in $\ket{0}$ state, by which we put the quantum system in an equal superposition of all possible solutions to the problem.

\noindent\textbf{Problem Hamiltonian.}
In this quantum system, each qubit represents a variable of the optimization problem. The problem Hamiltonian encodes the objective function of Eq.~\ref{eq:IsingObj} by mapping each variable $\variable{k}$ to a corresponding quantum spin $\spin{k}{z}$, which is the \textit{Pauli-z} gate shown in Table~\ref{table:gatetype}. We use $\proHam$ to represent the problem Hamiltonian, and it is known as an Ising Hamiltonian, as formulated below:
\begin{equation}
\proHam=\IsingObj{\spin{1}{z}, \spin{2}{z}, \ldots, \spin{\totalVariable}{z}}
\end{equation}
$\proHam$ is a matrix whose diagonal elements correspond to all the possible values of $\IsingObj{\binaryVariables}$. The ground state of $\proHam$ represents the solution to the optimization problem. To simulate the problem Hamiltonian with a quantum circuit, we need to leverage a unitary operator to be applied on the circuit, which is defined as:
\begin{equation}\label{eq:uniOp_problem}
\proUni{\gamma}=e^{-i\gamma\proHam}
\end{equation}
In this equation, $e$ is the natural logarithm, $i$ is the imaginary unit, and $\gamma$ represents the variational parameter of the QAOA algorithm, which determines the rotation amount of each qubit around the z-axis.

\noindent\textbf{Mixing Hamiltonian.}
The mixing Hamiltonian enables the algorithm to explore different parts of the search space and avoids getting stuck in local optima. It is typically chosen as:
\begin{equation}
\mixHam=\sum_{k=1}^{\totalVariable}\spin{k}{x}
\end{equation}
$\spin{k}{x}$ is the \textit{pauli-x} gate (shown in Table~\ref{table:gatetype}). A corresponding unitary operator is defined as:
\begin{equation}\label{eq:uniOp_mix}
\mixUni{\beta}=e^{-i\beta\mixHam}
\end{equation}
where $\beta$ is another variational parameter of the QAOA algorithm and determines the rotation amount of qubits around the x-axis.

The QAOA process is shown in Fig.~\ref{fig:qaoa}. The quantum circuit is first initialized with the Hadamard ($H$) gates. According to Table~\ref{table:gatetype}, each qubit will be in $\ket{+}$ state. Thus, we represent the whole circuit as in $\ket{+}^{\otimes \totalVariable}$ state.\footnote{For a quantum circuit with multiple qubits, we use the Kronecker product to describe the collective states.} We then apply the problem Hamiltonian and mixing Hamiltonian alternately $p$ times, which determines the depth of the QAOA algorithm \revision{(i.e., Fig.~\ref{fig:qaoa} shows a QAOA circuit of $p$ layers)}. Increasing the value of $p$ can theoretically improve the approximation quality and increase the required computational resources.
With the unitary operators defined in Eq.~\ref{eq:uniOp_problem} and Eq.~\ref{eq:uniOp_mix}, we calculate the quantum state as follows:
\begin{equation}\label{eq:wavefunction}
\begin{aligned}
\ket{\psi_p(\Vec{\gamma}, \Vec{\beta})}&= \proUni{\beta_p}\proUni{\gamma_p}, \ldots, \proUni{\beta_1}\proUni{\gamma_1}\ket{+}^{\otimes \totalVariable} \\
&= e^{-i\gamma_p\proHam}e^{-i\beta_p\mixHam}, \ldots, e^{-i\gamma_1\proHam}e^{-i\beta_1\mixHam}\ket{+}^{\otimes \totalVariable}
\end{aligned}
\end{equation}
Since the values of $\gamma$ and $\beta$ vary across different layers, the equation is defined with $2p$ parameters ($\Vec{\beta}=\beta_1, \ldots, \beta_p$ and $\Vec{\gamma}=\gamma_1, \ldots, \gamma_p$). The QAOA algorithm needs to find the optimal values of $\Vec{\beta}$ and $\Vec{\gamma}$ in order to find the minimum expectation value of $\proHam$, which corresponds to the optimal solution $\binaryVariables_{opt}$. The expectation value is calculated in the following:
\begin{equation}
\expEq{\Vec{\gamma}}{\Vec{\beta}}=\bra{\psi_p(\Vec{\gamma}, \Vec{\beta})}\proHam \ket{\psi_p(\Vec{\gamma}, \Vec{\beta})}
\end{equation}
The expectation value $\expEq{\Vec{\gamma}}{\Vec{\beta}}$ determines the results we measure from the quantum circuit. With such results, we can extract the solutions for $\IsingObj{\binaryVariables}$.
Specifically, the observed value of each qubit corresponds to the value of the variable $\variable{k}$ that it represents.\footnote{Taking the z-axis as the computational basis, if a qubit is observed as 0, the corresponding variable value is -1. Otherwise, the related variable value is +1.}.

In the QAOA algorithm, a classical optimizer is used to search for the optimal $\Vec{\gamma}$ and $\Vec{\beta}$ values. According to Fig.~\ref{fig:qaoa}, in each iteration, after the measurement, the classical optimizer updates the quantum circuit with the newly achieved parameters. The whole process repeats until a stopping condition is met.

\section{Methodology}\label{sec:methodology}
This section defines a generic Ising formulation for TCO problems (Sect.~\ref{subsec:ising}), illustrated with a running example followed by QAOA circuit (Sect.~\ref{subsec:QAOAcircuit}), and the process of \ourApproach (Sect.~\ref{subsec:overview}).

\subsection{Ising formulation for test case optimization}\label{subsec:ising}
We propose an Ising formulation for test case selection (TCS) and test case minimization (TCM) problems to be solved with QAOA.

\begin{definition}[{Test case selection (TCS)}]
Given a test suite $\testCaseSet =$ $\{\testCase{0},$ $\ldots, \testCase{\totalTestCase-1}\}$ with a set of testing attributes for the software under test and the corresponding testing objectives $\objSet=$ $\{\obj{0}, \ldots, \obj{\attrNum-1} \}$, TCS aims to select a subset $\seltestCaseSet \subset \testCaseSet$ to satisfy all objectives as much as possible. Each objective is associated with one of the attributes of the test cases. 
\end{definition}

\begin{definition}[{Test case minimization (TCM)}]
Given a test suite $\testCaseSet=\{ \testCase{0}, \ldots, \testCase{\totalTestCase-1}\}$ with a set of testing attributes for the software under test and the corresponding testing objectives, TCM aims to select minimum subset of test cases $\seltestCaseSet \subset \testCaseSet$ that satisfy all testing objectives as much as possible. The testing objectives $\objSet=\{ \obj{0},$ $\ldots,$ $\obj{\attrNum-1} \}$ include ones associated with attributes of the test cases and the one specific objective aiming to minimize the size of $\seltestCaseSet$, which makes TCM different from TCS.
\end{definition}

To construct the objective function of the TCO problem as an Ising formulation, we first represent $\variable{i}$ as the decision variable for selecting the test case $\testCase{i}$. $\variable{i}=-1$ denotes that $\testCase{i}$ is \textbf{selected} and $\variable{i}=+1$ that is \textbf{not selected}. Thus, the selection of all test cases is represented with a bit string $\binaryVariables \in \{-1, +1\}^{\totalTestCase}$.

TCO problems maximize the effectiveness of the selected test cases in test suite \seltestCaseSet, while minimizing the cost when executing the selected test cases. We divide all testing objectives into two groups according to their corresponding testing attributes: \textit{effectiveness objectives} and \textit{cost objectives}. 
Specially, for TCM problems, the objective of minimizing the size of selected test suite $\seltestCaseSet$ belongs to the \textit{cost objectives} group.
We minimize a cost-effective overall objective function for the TCO problem to seek the optimal solution $\binaryVariables^*$, which finds an optimal trade-off between these two attribute groups to find the approximate optimal solution that minimizes the corresponding Ising formulation.

We represent the objective function for TCO problems as $\objFunc(\obj{0},$ $\obj{1},$ $\ldots,$ $\obj{\attrNum-1})$. Since different attribute values are obtained with various metrics, we normalize all objectives ranging from 0 to 1 to ensure comparability.

Given the \textit{k}-th objective in the \textit{effectiveness objectives} group, we represent the corresponding attribute values of all test cases as $\effAttrValueSet{k}=\{\effValue{0}{k}, \ldots, \effValue{\totalTestCase-1}{k}\} $ and build the objective as follows: 
\begin{equation}\label{eq:effobj}
\obj{k}(\binaryVariables) = \frac{1}{2}\left(1+\frac{\sum_{i=0}^{\totalTestCase-1}\effValue{i}{k}\variable{i}}{\sum_{i=0}^{\totalTestCase-1}\effValue{i}{k}}\right), \quad \variable{i} \in \{+1,-1\}
\end{equation}
We normalize the range of the objective between 0 and 1. If a test case $\variable{i}$ is selected (i.e., $\variable{i}=-1$), the corresponding attribute value \effValue{i}{k} decreases, thus reducing the objective value. In contrast, if a test case is not selected (i.e., $\variable{i}=+1$), the attribute value is added to the objective value as a penalty.

Given the \textit{k}-th objective in the \textit{cost objectives} group, the corresponding attribute values of all test cases are represented as $\costAttrValueSet{k}=\{\costValue{0}{k}, \ldots, \costValue{\totalTestCase-1}{k}\}$. For TCM problems, with respect to the objective of minimizing the size of the selected test suite, the attribute value for each test case is assigned as 1. We build the corresponding objective in the following:
\begin{equation}\label{eq:costobj}
\obj{k}(\binaryVariables) = \frac{1}{2}\left(1-\frac{\sum_{i=0}^{\totalTestCase-1}\costValue{i}{k}\variable{i}}{\sum_{i=0}^{\totalTestCase-1}\costValue{i}{k}}\right),\quad \variable{i} \in \{+1,-1\} 
\end{equation}
The value of the objective is still in the range of 0 and 1. If a test case $\variable{i}$ is selected ($\variable{i}=-1$), the corresponding attribute value \costValue{i}{k} will be added to the objective value as a penalty since a higher number of selected test cases will increase the cost; otherwise, the attribute value will contribute towards decreasing the objective value.

For each objective, to emphasize the larger deviation between the obtained value and the theoretical optimal value (i.e., 0), we square all objectives when combined to amplify the impact and penalize the improper selection.
For TCO problems, we formulate the overall objective function $\objFunc(\obj{0}, \obj{1}, \ldots, \obj{\attrNum-1})$ as:
\begin{equation}
\min{\objFunc_{TCO}(\binaryVariables) = \weight{0}(\obj{0})^2+\ldots+\weight{\attrNum-1}(\obj{\attrNum-1})^2}
\end{equation}
$\weight{0}, \weight{1}, \ldots, \weight{\attrNum-1}$ are weights assigned to various objectives\revision{, indicating the degrees of importance of each objective, based on user requirements in practice. These weights, which sum up to 1, $\sum_{i=0}^\attrNum \weight{i}=1$, can be freely chosen based on user preferences, and in some cases, domain experts can provide the weight values. While the selection of weights changes the optimal solution of the problem, the performance of \ourApproach is not dependent on the weights selected.} 


Suppose there are \numEffAttr objectives in the \textit{effectiveness objectives} group and \numCostAttr objectives in the \textit{cost objectives} group ($\numEffAttr+\numCostAttr=\attrNum$). Since $\variable{i}\in\{-1,+1\}$, it holds $\variable{i}^2=1$. We expand the objective function as:
\begin{equation}\label{eq:final_obj}
\objFunc_{TCS}(\binaryVariables) = \sum_{\langle i, j\rangle}X_{ij}\variable{i}\variable{j} + \sum_iY_{i}\variable{i} + \const
\end{equation}
\begin{equation*}
X_{ij} = \sum_{k=0}^{\numEffAttr-1}\weight{k}\frac{\effValue{i}{k}\effValue{j}{k}}{\left(\sum_{l=0}^{\totalTestCase-1}\effValue{l}{k}\right)^2} - \sum_{k=0}^{q-1}\weight{k+\numEffAttr}\frac{\costValue{i}{k}\costValue{j}{k}}{\left(\sum_{l=0}^{\totalTestCase-1}\costValue{l}{k}\right)^2}
\end{equation*}
\begin{equation*}
Y_i = \sum_{k=0}^{\numEffAttr-1}\weight{k}\frac{\effValue{i}{k}}{\sum_{j=0}^{\totalTestCase-1}{\effValue{j}{k}}} - \sum_{k=0}^{q-1}w_{k+p}\frac{\costValue{i}{k}}{\sum_{j=0}^{n-1}\costValue{j}{k}}
\end{equation*}

\subsection{Realizing TCO in a QAOA Circuit} \label{subsec:QAOAcircuit}
From the objective function in Eq.~\ref{eq:final_obj}, we first remove \const because it does not affect the optimization process and only changes the absolute value of the fitness value. Then, we replace each variable $\variable{j}$ with $\spin{j}{z}$ to create the corresponding problem Hamiltonian.
\begin{equation}
H_C = \sum_{\langle j, k \rangle}X_{j,k}\spin{j}{z}\spin{k}{z} + \sum_jY_{j}\spin{j}{z}
\end{equation}
When creating the quantum circuit, we represent each test case with one qubit. $\spin{j}{z}$ corresponds to placing a \textit{Pauli-z} gate on the $j$-th qubit, which is explained in Sect.~\ref{sec:background}.
To simulate the problem Hamiltonian in a quantum circuit, we need to create the unitary operator:
\begin{equation}
\proUni{\gamma}=e^{-i\gamma\proHam}=e^{-i\gamma \sum_{\langle j, k \rangle}X_{j,k}\spin{j}{z}\spin{k}{z}}e^{-i\gamma \sum_j^{\totalTestCase}Y_{j}\spin{j}{z}}
\end{equation}

\begin{figure}[!tb]
\centering
\includegraphics[width=\columnwidth]{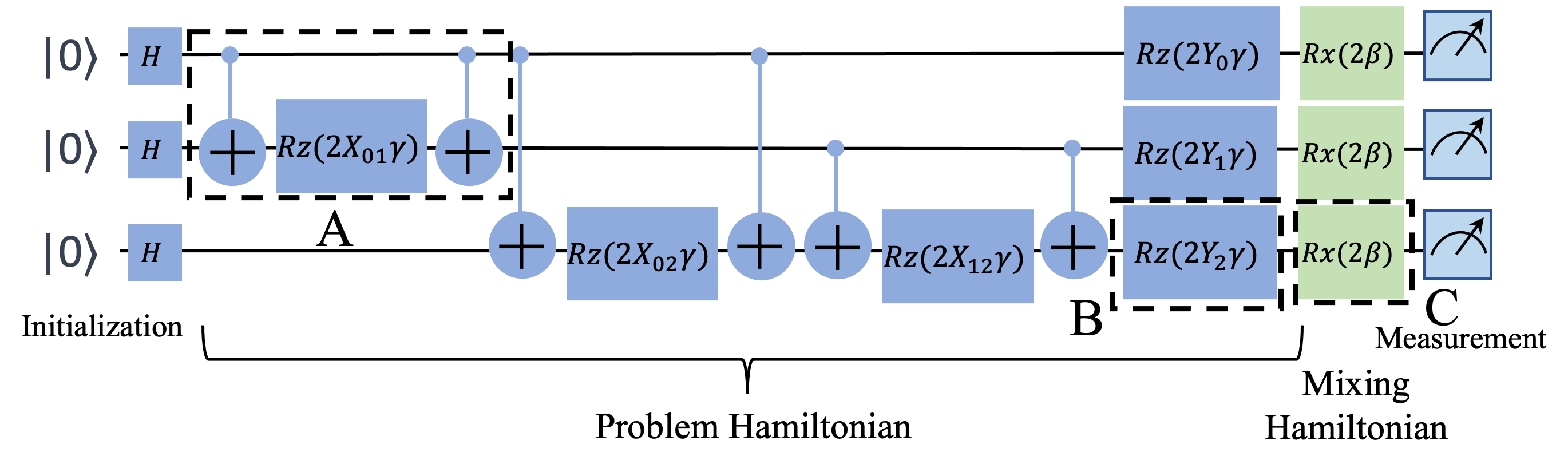}
\caption{QAOA example circuit. A typical 1-layer circuit from the literature is used for TCO problems with 3 test cases. Each test case corresponds to one qubit}
\label{fig:circuit_example}
\end{figure}

According to~\cite{whitfield2011simulation}, to create the unitary operator $e^{-j\gamma Y_{j}\spin{j}{z}}$ related to the $j$-th test case (i.e., the $j$-th qubit), we use the $\mathit{Rz}(\theta)$ defined in Table~\ref{table:gatetype} in Sect.~\ref{sec:background}. In a quantum circuit, we can apply the gate on the $j$-th qubit with rotation $\theta = 2Y_j \gamma$ around the z-axis where \revision{$Y_j$} is the corresponding coefficient of the linear term in the objective function, and $\gamma$ is the parameter of problem Hamiltonian to optimize (introduced in Sect.~\ref{sec:background}). For each qubit, if the corresponding linear coefficient $Y_{j}$ is different from $0$, we apply gates inside the box labeled $B$ in Fig.~\ref{fig:circuit_example} on the $j$-th qubit.

For the unitary operator $e^{-i\gamma X_{j,k}\spin{j}{z}\spin{k}{z}}$ related to the $j$-th and $k$-th test cases (i.e., the $j$-th and $k$-th qubits), according to~\cite{whitfield2011simulation}, we can use the $\mathit{CNOT}$ gate first to entangle two qubits, then apply the $Rz(\theta)$ gate followed by another $\mathit{CNOT}$ gate, which creates the unitary gate $e^{-i\spin{j}{z}\spin{k}{z}\theta/2}$. We depict this gate combination inside the box labeled $A$ in Fig.~\ref{fig:circuit_example}. The angle parameter of $Rz(\theta)$ gate is $\theta = 2X_{\mathit{jk}} \gamma$, where $X_{\mathit{jk}}$ is the coefficient of the corresponding interaction term. For every pair of qubits, if the corresponding coefficient $X_{\mathit{jk}} \neq 0$, we apply the set of gates to the two qubits.

Similarly, for the mixing Hamiltonian,
\begin{equation}
\mixHam=\sum_{j=1}^{\totalVariable}\spin{j}{x}, \quad \proUni{\beta}=e^{-i\beta\mixHam} = e^{-i\beta\sum_{j=1}^{\totalVariable}\spin{j}{x}}
\end{equation}
Since $Rx(\theta)$ generates unitary operator of $e^{-i\theta/2\spin{j}{x}}$, which rotates $\theta$ angle around the x-axis (see Sect.~\ref{sec:background}), the angle parameter should be $\theta = 2\beta$ for each qubit to create the unitary operator $e^{-i\beta\spin{j}{x}}$, where $\beta$ is the parameter of the mixing Hamiltonian to be optimized (introduced in Sect.~\ref{sec:background}). Thus, for the mixing Hamiltonian, we apply a gate of the box labeled $C$ on each qubit as shown in Fig.~\ref{fig:circuit_example}.

\begin{example} 
We demonstrate a QAOA circuit with depth $p=1$ using a small-scale running example for a TCM problem to construct an objective function. Assume we have three test cases \testCase{0}, \testCase{1}, and \testCase{2}, with their \textit{failure rate} values of 50\%, 70\%, and 80\%, respectively, and \textit{execution time} of 3, 6, and 1 second, respectively. The corresponding objective of the \textit{failure rate} is an \textit{effectiveness objective}, and the objective for \textit{execution time} is a \textit{cost objective}. Together with the objective of minimizing the number of selected test cases, we assign equal weights (i.e., 1/3) \revision{as an example} to the three objectives. \revision{Note that any other weights can be selected depending on the user's preference.} \revision{We} build the objective function below according to Eq.~\ref{eq:final_obj}:
\begin{equation}
\begin{aligned}
\mathcal{O}_{TCM}(\mathbf{z}) &= X_{01}z_{0}z_{1} + X_{02}z_{0}z_{2} + X_{12}z_{1}z_{2} \\
&+Y_0z_{0}+Y_1z_{1}+Y_2z_{2} + \const
\end{aligned}
\end{equation}
\begin{equation}
\begin{aligned}
&X_{01}=0.063, X_{02}=0.040, X_{12}=0.052,\\
&Y_{0}=-0.064,Y_{1}=-0.097,Y_{2}=-0.0069,
\end{aligned}
\end{equation}


The quantum circuit is shown in Fig.~\ref{fig:circuit_example}. 
Three test cases are mapped to 3 qubits. We first initialize all qubits in $\ket{0}$ state with the Hadamard gate. Next, we create the circuit for the problem Hamiltonian. Because the coefficients of all interaction terms are not 0, we need to entangle all three qubits with each other using the gates set in box $A$. Next, regarding all linear terms, we apply the $\mathit{Rz}$ gate on each qubit. Then, the mixing Hamiltonian is created with the $\mathit{Rx}$ gate on each qubit. Since $p=1$, we directly apply measurements on each qubit to read the problem's possible solution. If $p>1$, the problem Hamiltonian and mixing Hamiltonian should be applied $p$ times alternately before the measurement.
\end{example}
\subsection{Process of \ourApproach}\label{subsec:overview}
\begin{figure}
\centering
\includegraphics[width=\columnwidth]{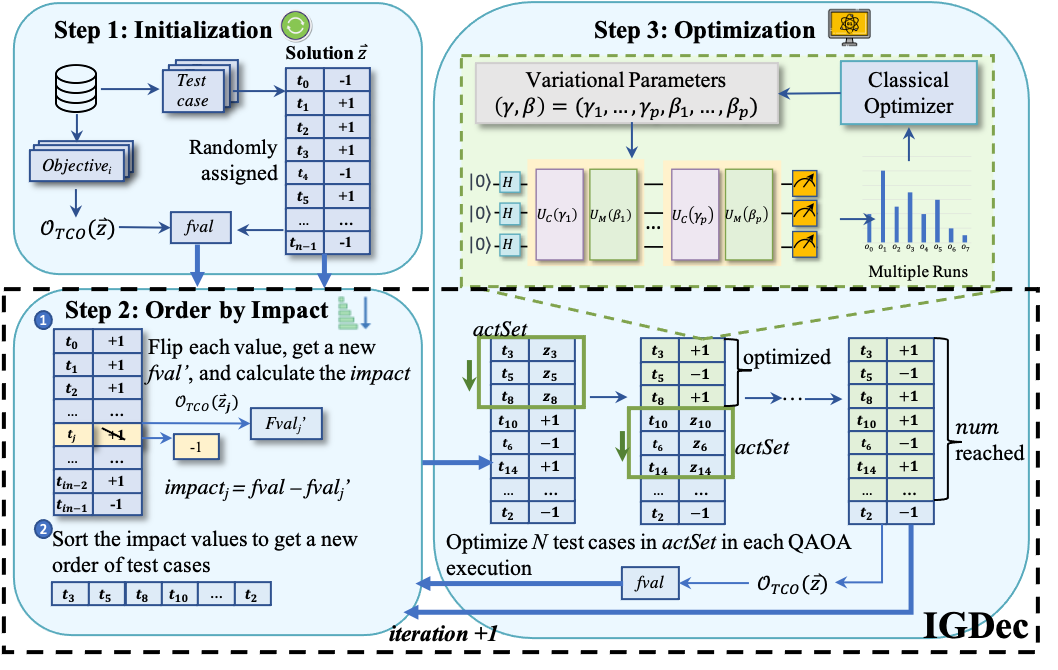}
\caption{Process of \ourApproach. It has three steps: initialization, test case ordering by impact, and optimization}
\label{fig:overview}
\end{figure}

Currently, available quantum computers with limited qubits cannot directly solve TCO problems with many test cases with QAOA. Thus, we combine an impact-guided decomposition (IGDec) strategy adapted from \textit{qbsolv}~\cite{dwavePartitioning} by D-Wave with QAOA to solve TCO problems as a classical-quantum hybrid approach with the Ising formulation and QAOA circuit described before. \revision{In the IGDec strategy, the decomposition of the problem is determined by the \textit{impact} value of each test case.}
Fig.~\ref{fig:overview} shows the overview of \ourApproach that has the three main steps described below, \revision{and the IGDec strategy is composed of Step 2 and Step 3.}

\textbf{Step 1. Initialization.}
Given a test suite containing \totalTestCase test cases, we initially assign random values (i.e., $-1$ or $+1$) to the decision variables associated with each test case. We represent the bit string as $\binaryVariables$. With the objective function of Eq.~\ref{eq:final_obj}, we get an initial fitness value $\mathit{fval}$.

\textbf{Step 2. Ordering test cases by \textit{impact}.}
\revision{This step constitutes the ordering phase of the IGDec strategy. We first calculate the \textit{impact} value of each test case, and then sort the test cases according to this value.} This impact value indicates how much a test case contributes to reducing the fitness value of the TCO problem when its decision variable is altered. We use the calculation of the impact value of \testCase{k} as an example. Taking the bit string of the solution $\binaryVariables$ and the associated fitness value $\mathit{fval}$, for a test case \testCase{k}, we flip the value of its decision variable $\variable{k}$ (i.e., $-1$ to $+1$ or $+1$ to $-1$) to get a new bit string and the corresponding new fitness value $\mathit{fval}_{k}'$. Next, we calculate the difference between $\mathit{fval}$ and $\mathit{fval}_{k}'$, which is the $\mathit{impact}$ of \testCase{k} (i.e., $\mathit{impact}_{k} = \mathit{fval}_{k}' - \mathit{fval}$). In this way, we calculate the impact values of each test case and sort them based on the values.
Since we are solving a minimization problem, we arrange the test cases in ascending order of impact values. The test cases with lower impact values are positioned towards the front because they significantly decrease the fitness value when flipped.

\textbf{Step 3. Optimization.}
\revision{This step constitutes the decomposition phase of the IGDec strategy.} With the rearranged test cases, we decompose the problem and run QAOA iteratively to get the approximate optimal solution. We first determine the sub-problem size \subproblemSize that a quantum computer can afford. During the optimization process, we always optimize over a set of \subproblemSize variables in each QAOA execution, which belong to the \textit{active} variable set \activeSet. For other variables, the values are fixed, maintaining the current values. Thus, with the overall objective function from Eq.~\ref{eq:final_obj}, irrespective of the constant term, we get the objective function:
\begin{equation}\label{eq:clamp}
\objFunc_{TCO}(\binaryVariables) = \sum_{i,j\in \activeSet}X_{ij}\variable{i}\variable{j} + \sum_{i\in \activeSet}(Y_i+\sum_{j\notin\activeSet}X_j\variable{j}^*)\variable{i} 
\end{equation}
$\variable{j}^*$ refers to the fixed values of variables not in set \activeSet. We optimize the variables (i.e., test cases) in order.
In the initial execution, we select the first \subproblemSize variables as the \textit{active} variables. Then, we run the QAOA algorithm on the objective function to get optimized values of test cases in the \textit{active} variables. Each variable in \activeSet corresponds to one qubit. Next, we optimize the next \subproblemSize variables, with the first \subproblemSize variables keeping the current optimized values. We continuously run the QAOA algorithm until \fractionNum test case variables are optimized. With the new bit string $\binaryVariables$, we get the new fitness value $\mathit{fval}$. If the stopping condition is not met, we return to \textbf{Step 2} and start a new iteration with the new bit string $\binaryVariables$ and $\mathit{fval}$, then arrange the test cases again.

Throughout the process of \ourApproach, we identify and mark the best $\mathit{fval}$ (i.e., the lowest fitness value) and the corresponding bit string $\binaryVariables$ (i.e., the solution). We repeat \textbf{Step 2} and \textbf{Step 3} until the stopping condition is achieved. Upon reaching this condition, the marked $\binaryVariables$ \revision{is} recognized as the optimal solution to the problem.

\section{Experiment Design}\label{sec:expDesign}

\subsection{Research Questions} \label{subsec:RQs}

\begin{compactitem}
\item[\textbf{RQ1}]
How do the number of layers and the sub-problem size affect the performance of \ourApproach?
\item[\textbf{RQ2}] How is the performance of \ourApproach compared with that of the baseline QAOA approach?
\item[\textbf{RQ3}] How is the performance of \ourApproach compared with that of classical approaches?
\item[\textbf{RQ4}] How is the performance of \ourApproach on a noisy quantum computer?
\end{compactitem}

RQ1 studies how the two key parameters of \ourApproach affect its performance. The first parameter is the depth of the QAOA circuit (i.e., the number of layers \textit{p}). A higher value of \textit{p} incurs higher cost in terms of execution. It is, therefore, important to study which number of layers is optimal for TCO problems. The second parameter is the sub-problem size (i.e., $N$) to be solved on QAOA and how a particular sub-problem size affects the performance of \ourApproach. This aspect helps us study which sub-problem size is optimal for TCO. In addition, we study whether a higher number of \subproblemSize incurs more execution time, i.e., studying the cost associated with \subproblemSize. 

RQ2 studies whether \ourApproach's test case ordering by the IGDec strategy helps the QAOA to find approximately optimal solutions. To this end, we replace \ourApproach's IGDec strategy with a random strategy while keeping the QAOA part the same. 

RQ3 compares \ourApproach with two classical search algorithms: Random Search (RS) and a Genetic Algorithm (GA). We assess whether we can obtain optimal solutions close to, or even outperform, the classical search algorithm, i.e., GA, whereas comparison with RS ensures that we solve a complex problem that cannot be solved randomly.

\revision{RQ4 analyzes the effect of noise on \ourApproach. We run QAOA on a noisy simulator and compare the results with those obtained by \ourApproach on an ideal simulator, as well as those obtained by GA. Moreover, we run a small case study on IBM's quantum computer, which aims to demonstrate the feasibility of our approach on a real quantum computer in the presence of noise.}

\subsection{Experiment Setup} \label{subsec:experimentsetup}
This section discusses datasets we used for experiments, baselines, parameter settings, and execution environment.

\subsubsection{Datasets}
We employed four publicly available industrial datasets. 

The \textit{Paint Control} dataset contains 90 test cases, and the \textit{IOF/ROL} dataset has 1941 test cases. Both of them are from ABB Robotics Norway~\cite{spieker2017reinforcement}. The \textit{GSDTSR} dataset\footnote{\url{https://code.google.com/archive/p/google-shared-dataset-of-test-suite-results/wikis/DataFields.wiki}} from Google has 5555 test cases. Each test case contains two attributes: "execution time" and "failure rate". The first attribute contains a value telling how much time it was taken on average by a test case to execute, whereas the failure rate determines the possibility of a test case failing based on the historical ratio of failures to the total number of test case executions.
We run experiments on these datasets for the TCM problems. To do so, we encode the objective of minimizing the number of selected test cases. The overall fitness function objectives are to minimize the test execution cost and the number of selected test cases while maximizing the fault detection rate of selected test cases. 

The \textit{ELEVATOR} dataset is from Orona\footnote{\url{https://www.orona-group.com/int-en/}}~\cite{9978988}, which was produced when testing a dispatcher, a software component in elevators responsible for efficient elevator scheduling while maximizing service quality (e.g., reducing passengers' waiting time) in the software in the loop setup. The dataset has a total of 1925 test cases, and we selected two fitness functions to form the two TCS problems:
\begin{inparaenum}[1)]
\item ELEVATOR$_{o2}$: Minimizing the overall test case execution of selected test cases (i.e., cost reduction) while maximizing the test case diversity (called ``input diversity'').
\item ELEVATOR$_{o3}$: Minimizing the overall test case execution of selected test cases (i.e., cost reduction) while maximizing passenger counts and travel distance in the elevator system. 
\end{inparaenum}

\subsubsection{Baselines}\label{sec:baseline}

We selected the following baselines for comparison with \ourApproach. 

\noindent \textbf{\baselineqaoa.} We implemented as a baseline approach a version of \ourApproach without the IGDec strategy. The aim is to assess whether the IGDec strategy contributes to the overall performance of \ourApproach in solving TCO problems. Specifically, we start with \textbf{Step 1} of QAOA from Sect.~\ref{subsec:overview} to randomly initialize variable values for each test case with -1 or +1. Next, we randomly select test cases as \textit{active variables} into the \activeSet, whose size is fixed. Similarly to \textit{Step 3} in Sect.~\ref{subsec:overview}, with one \activeSet, we use Eq.~\ref{eq:clamp} to optimize sub-problems. We randomly select the second \activeSet with the newly optimized variable values. We repeatedly select and run sub-problems until a determined number of QAOA runs has been reached. During the optimizing process, we ensure each variable is selected as the \textit{active variables} and optimized at least once. We record the fitness values achieved after each run. Finally, the solution after the last sub-problem is solved is considered the final solution.

\noindent \textbf{Genetic Algorithm.} We used the implementation of the Genetic Algorithm (GA) provided by the jMetalPy 1.6.0~\cite{benitez2019jmetalpy} framework. We represented each individual of the TCO problems as a bit string. We used the default parameter settings from jMetalPy, i.e., bit flip mutation operation being equal to the reciprocal of the size of the individual, the SPX crossover (the crossover rate=1.0). For each dataset, we experimented with population size $\pop=\{10,$ $20,$ $30,$ $40,$ $50,$ $60,$ $70,$ $80,$ $90,$ $100\}$. We ran 400,000 evaluations for each experiment to ensure that GA had sufficient time to converge. Note that in the literature, the default parameter settings of GA have shown good results~\cite{ParameterTunning}. 

\noindent \textbf{Random Search.} We used the existing random search (RS) implemented by jMetalPy 1.6.0~\cite{benitez2019jmetalpy}. It randomly selects binary values for each test case and calculates the fitness value repeatedly until a determined number of iterations has been reached, and we pick the lowest fitness value to compare with other approaches.


\subsubsection{Parameters and Execution Environment} \label{subsubsec:parameters} 

In \textbf{Step 2} of \ourApproach, we set $\fractionNum=0.15\times\totalTestCase$ (see~\cite{dwavePartitioning}) for larger datasets (i.e., \textit{IOF/ROL}, \textit{GSDTSR}, and \textit{ELEVATOR}). For the smaller dataset (i.e., Paint Control), if \fractionNum is smaller than the sub-problem size \subproblemSize, we set $\fractionNum=\subproblemSize$ to ensure that at least one execution of QAOA in each iteration takes place. \ourApproach has the termination condition of both \textbf{Step 2} and \textbf{Step 3} (as detailed in Sect.~\ref{subsec:overview}), i.e., no reduction in $\mathit{fval}$ for three consecutive iterations. \revision{The maximum generation is set as 30.} We run \ourApproach with $\subproblemSize=\{7,$ $8,$ $10,$ $12,$ $14,$ $16\}$ considering the execution time limitation. For all experiments, including \ourApproach and the three baseline approaches, we repeat \revision{them 30} times \revision{to account for the} randomness inherent in the approaches.
We use \textit{aer simulator} in Qiskit to run the QAOA algorithm. \revision{For the noisy quantum computer, we implement QAOA on IBM's quantum computer ``ibm\_brisbane'' and a simulator equipped with a noise model based on this quantum computer.} As a classical optimizer, we select the \textit{Constrained Optimization By Linear Approximation} optimizer (COBYLA)~\cite{powell2007view}, with 100 iterations, which is commonly used in literature for the QAOA algorithm. We run QAOA with $p=\{1,$ $2,$ $4,$ $8,$ $16\}$. The experiment was run on one CPU node in the eX3\footnote{eX3 is a national, experimental, heterogeneous computational cluster for researchers conducting experiments.} cluster with AMD EPYC 7601 32-Core Processor.

\subsection{Evaluation metrics and statistical tests}

\subsubsection{Evaluation metrics}
Approximation ratio (\appro) is a common evaluation metric for assessing QAOA's effectiveness in finding approximate optimal solutions~\cite{QAOAPower}. We calculate \appro as:
\begin{equation}
\appro = \frac{\mathit{fval}(\binaryVariables)}{\mathit{fval_{min}}}
\end{equation}
\noindent where $\mathit{fval_{min}}$ is the optimal solution of the problem and $\mathit{fval}(\binaryVariables)$ is the final solution obtained by QAOA. We use the lowest fitness value achieved in the whole experiment as the optimal fitness value for each case study. For \appro, a lower value is indicative of a better performance. The optimal \appro is 1.0 for all case studies.

To evaluate the time efficiency, we calculate the execution time \extime. For \ourApproach, it contains both the QAOA execution time and classical computation time, including the time for the IGDec strategy.

In addition, we define fitness evaluations \numEval to help analyze the performance of each approach.
For \ourApproach, we calculate the number of QAOA runs (i.e., the sub-problems solved by QAOA) in all iterations until the stopping condition is satisfied as \numEval. Similarly, for \baselineqaoa, the number of randomly selected sub-problems is considered as \numEval. For GA, \numEval refers to the number of evaluations through all generations. For RS, each run (i.e., iteration) represents one evaluation, and the determined number of iterations represents the \numEval value (each iteration contains one run).

\subsubsection{Statistical tests}
Following the guideline from~\cite{arcuri2011practical}, we apply statistical tests to evaluate the performance (i.e., \appro) of \ourApproach with different layers (i.e., $p$) and sub-problem sizes (i.e., \subproblemSize), and the baseline approaches. We first utilize a non-parametric Kruskal-Wallis H-test to check whether there is a statistical difference among all groups of \appro values obtained by \ourApproach with different configurations or baseline approaches. We set the significance level as 0.05. A p-value smaller than 0.05 indicates that at least two groups of \appro values are significantly different.

Next, we apply the Mann-Whitney U test and Vargha and Delaney's \Atwelve effect size to compare all possible pairs of \ourApproach with different configurations, and we also compare \ourApproach with the other baseline approaches respectively. We set the significance level as 0.05. A p-value less than 0.05 indicates that there is a significant difference between the \appro values of the two selected groups. In this case, we utilize Vargha and Delaney's \Atwelve statistic as the effect size measure to quantify the magnitude of the difference between the two groups. If \Atwelve is 0.5, the result is achieved by chance. If \Atwelve is smaller than 0.5, \revision{the first approach performs better than the second one; otherwise, the second one is better.}

\section{Results and analyses}\label{sec:expResults}
This section presents the results and analyses for each RQ, followed by the threats to validity. 

\subsection{RQ1 -- Selecting Optimal Depth and Sub-problem Size}\label{subsec:RQ1results}
As discussed in Sect.~\ref{subsec:RQs}, we study two key parameters of \ourApproach, i.e., depth \textit{p} and sub-problem size $\subproblemSize$. 

\noindent\textbf{Results for Optimal Depth.} To study whether the depth of a QAOA circuit affects the performance of \ourApproach, we compared the approximate values for \revision{30} runs of each \textit{p} value (i.e., 1, 2, 4, 8, and 16, Sect.~\ref{subsubsec:parameters}) for all sub-problem sizes (i.e., 7, 8, 10, 12, 14, and 16 in Sect.~\ref{subsubsec:parameters}) for each case study. As a result, for each \textit{p} value, when combining all sub-problem results, we obtain \revision{180} approximation ratio values (i.e., \appro). To assess the overall differences among all the \textit{p} values, we applied the Kruskal-Wallis H-test. We obtained p-values greater than 0.05, indicating no significant differences among the number of layers chosen for QAOA solving the test optimization problems for all the chosen case studies. Using $p=1$ in our test optimization problem makes sense since it is the least costly. Thus, to study the effect of sub-problem size in this RQ, we chose $p=1$. 



\begin{figure*}[!tb]
\centering
\begin{subfigure}{.195\textwidth}
\centering
\includegraphics[width=1\linewidth]{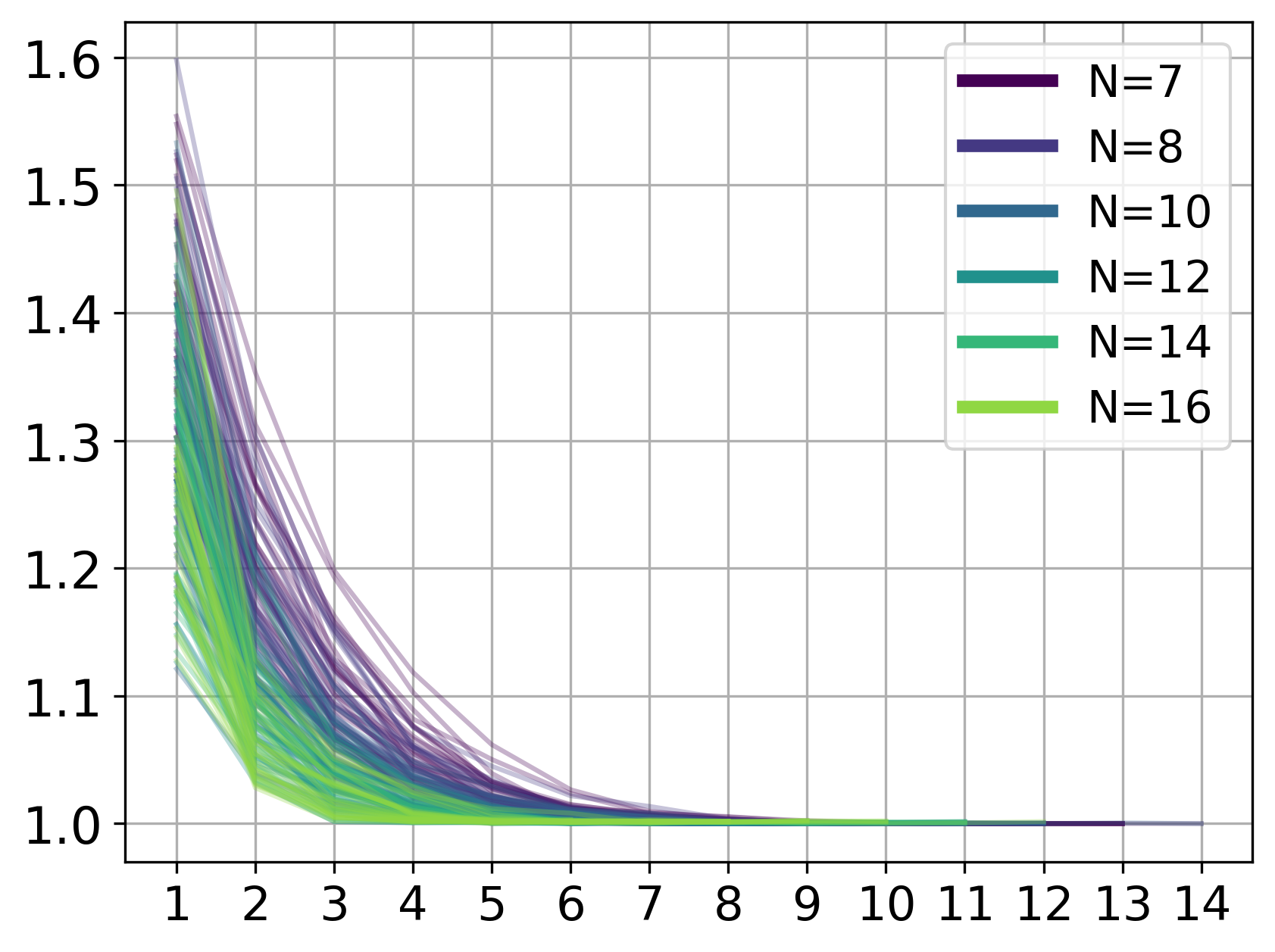}
\caption{\revision{Paint Control}}
\label{fig:trendPaint}
\end{subfigure}%
\begin{subfigure}{.195\textwidth}
\centering
\includegraphics[width=1\linewidth]{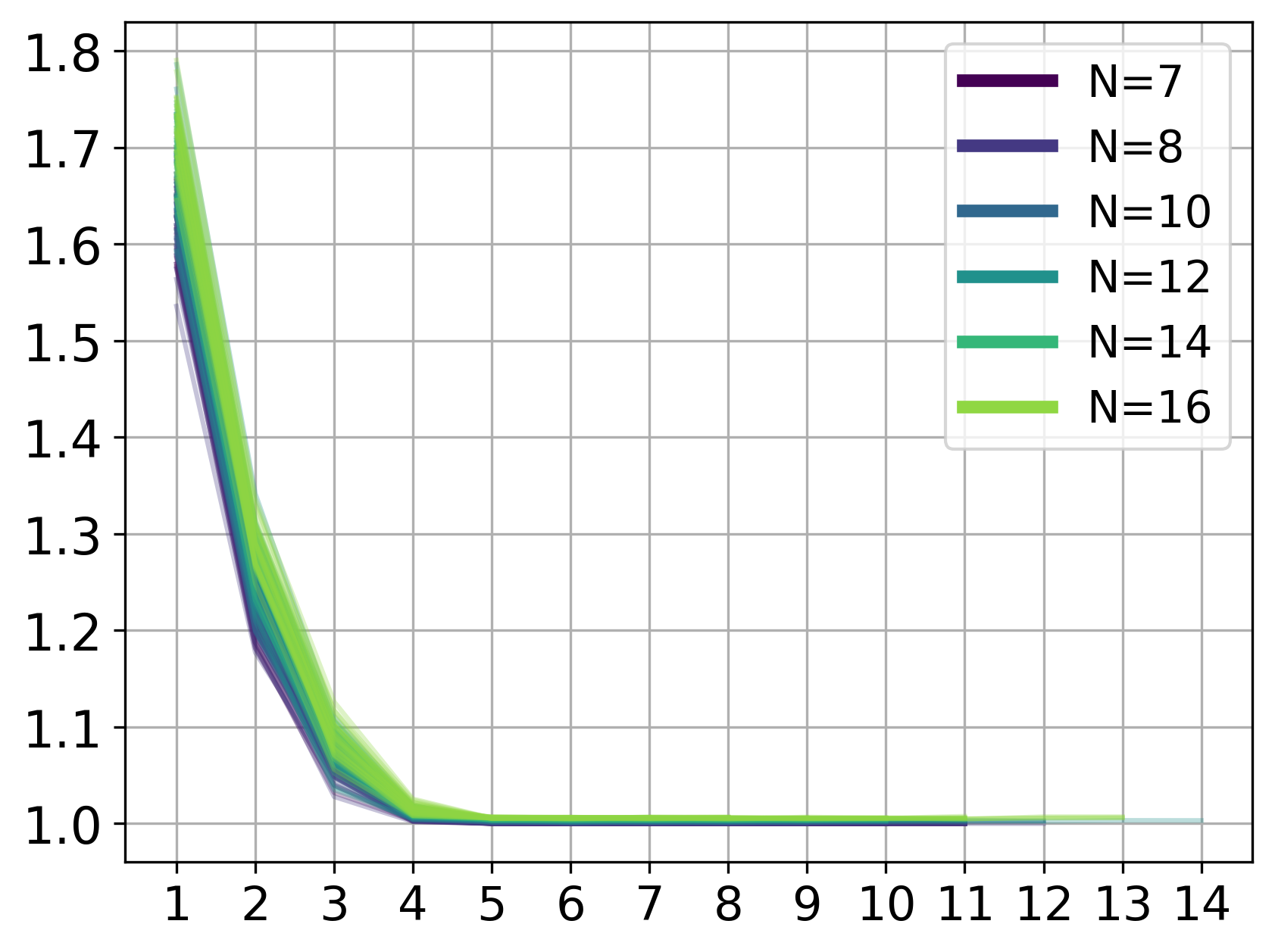}
\caption{\revision{IOF/ROL}}
\label{fig:trendIOF}
\end{subfigure}
\begin{subfigure}{.195\textwidth}
\centering
\includegraphics[width=1\linewidth]{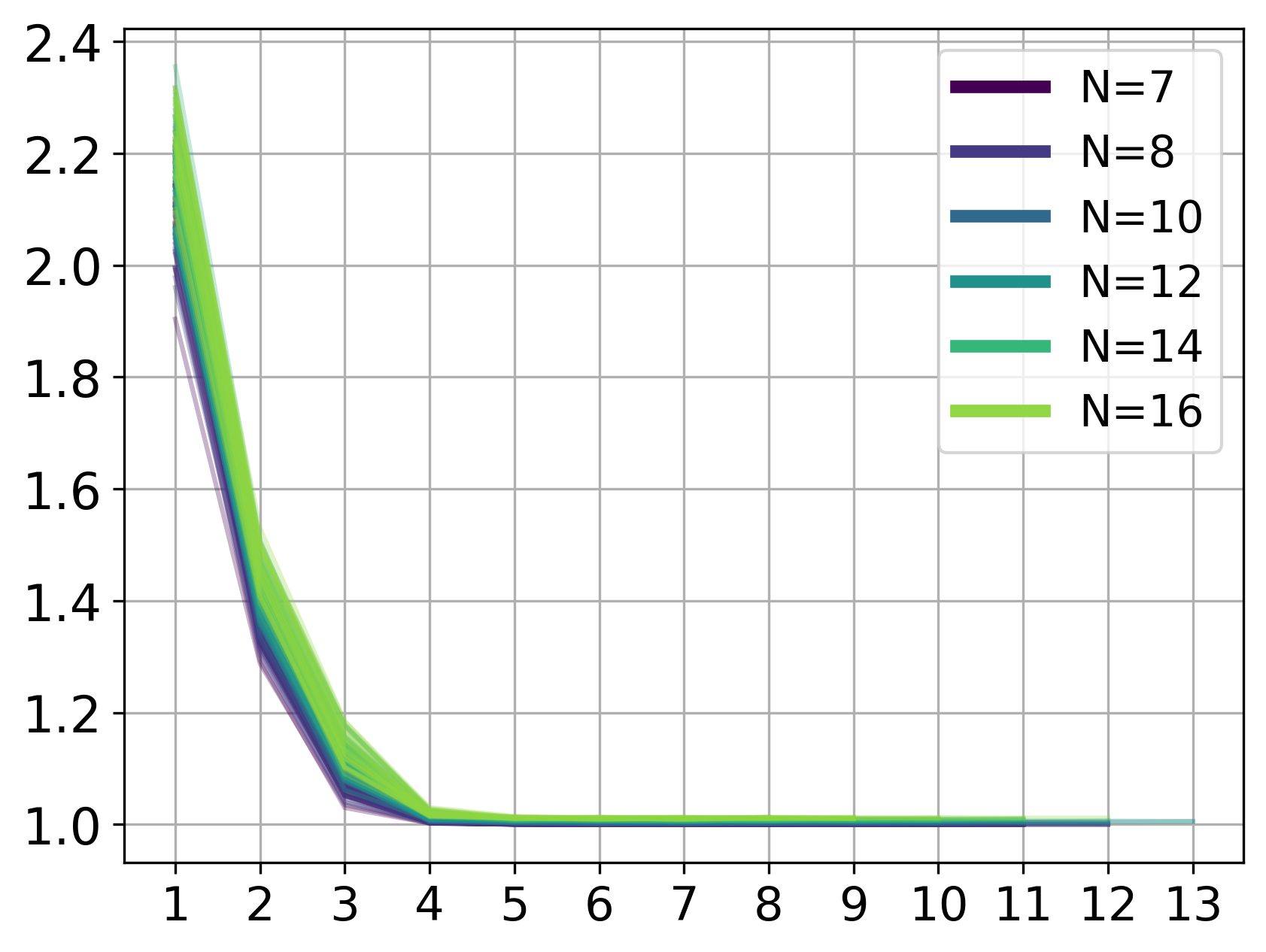}
\caption{\revision{ELEVATOR$_{o2}$}}
\label{fig:trendELEV02}
\end{subfigure}
\begin{subfigure}{.195\textwidth}
\centering
\includegraphics[width=1\linewidth]{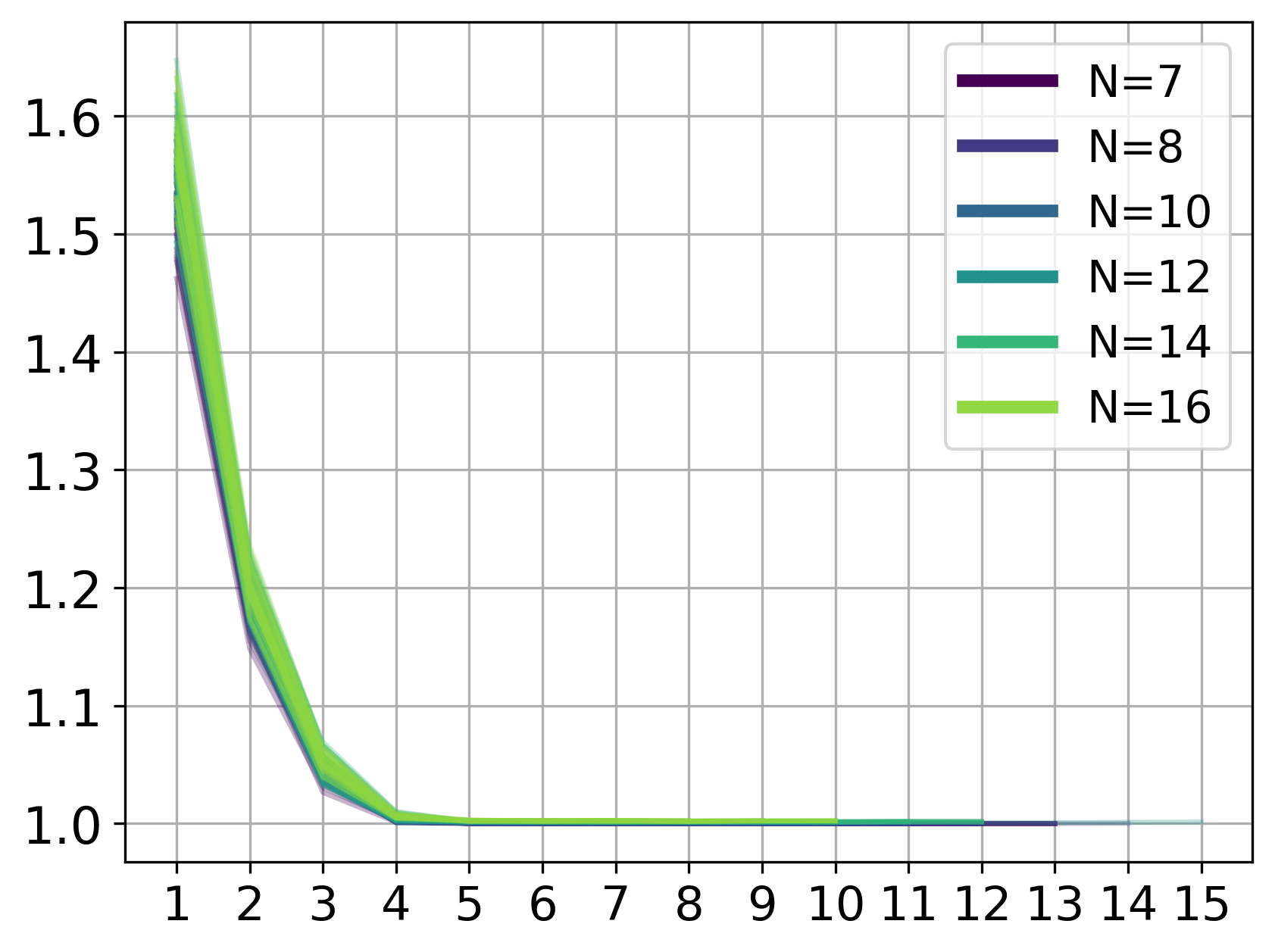}
\caption{\revision{ELEVATOR$_{o3}$}}
\label{fig:trendELEV03}
\end{subfigure}
\begin{subfigure}{.195\textwidth}
\centering
\includegraphics[width=1\linewidth]{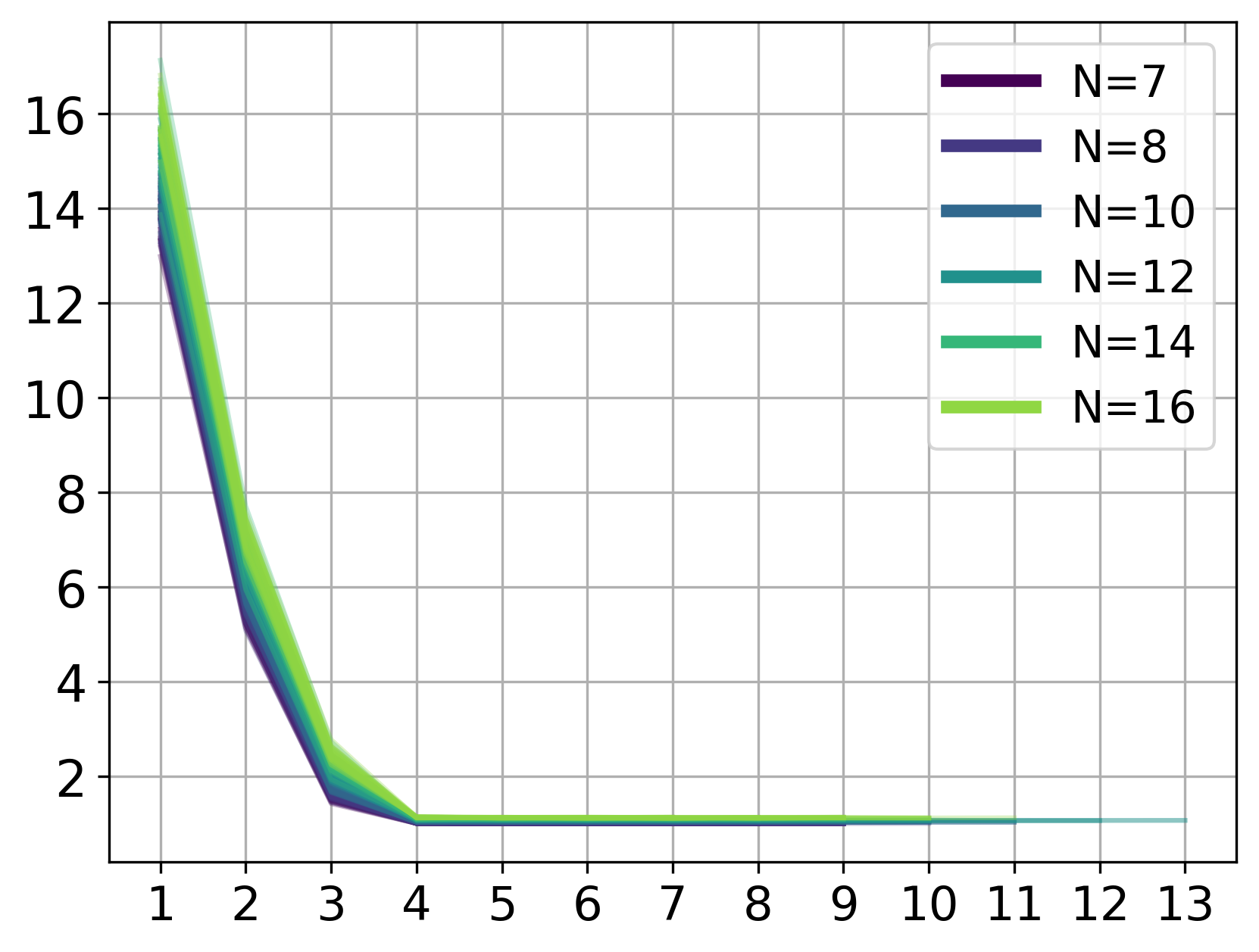}
\caption{\revision{GSDTSR}}
\label{fig:trendGSDTSR}
\end{subfigure}
\caption{\revision{RQ1 -- Trends of approximation ratios (\appro) along the iterations of \ourApproach. The x-axis shows the iteration number, whereas the y-axis shows the approximation ratio. For each sub-problem size (i.e., \subproblemSize = 7, 8, 10, 12, 14, or 16), we show \revision{30} trend lines corresponding to \revision{30 runs}.} }
\label{fig:trend}
\end{figure*}

\noindent\textbf{Results for Optimal Sub-Problem Size.} We also study whether there is an optimal sub-problem size for \ourApproach solving the test optimization problems. We present the results for all five case studies in Figs.~\ref{fig:trendPaint}-\ref{fig:trendGSDTSR}.
Each figure shows the trend of approximation ratio for \revision{30} runs of each sub-problem size along the increasing iterations, i.e., we see \revision{30 trend} lines for each sub-problem size. 
One observation is that the \appro values decrease dramatically in the first three iterations, consistently for all case studies, and stabilize to near 1.0 (the best \appro). Note that for all our case studies, except \textit{Paint Control}, \ourApproach with smaller $N$ (darker purple lines) tends to achieve lower \appro in each iteration, which indicates a faster decline rate. However, the results are a bit different for \textit{Paint Control}, for which the values of \appro achieved by \ourApproach with larger sub-problem sizes tend to decline more quickly. As the dataset size of \textit{Paint Control} is small (i.e., only 90 test cases) and the number of optimized test cases cannot exceed $0.15\times \totalTestCase$ (Sect.~\ref{sec:expDesign}), for smaller \subproblemSize (i.e., 7, 8, 10, and 12), QAOA can only run once in each iteration. However, for larger \subproblemSize (i.e., 14 and 16), we also need to ensure at least one run of QAOA is executed in each iteration. This indicates that \fractionNum for larger \subproblemSize is greater than that for small \subproblemSize. Thus, \ourApproach with larger \subproblemSize can optimize more test cases in each iteration such that the optimization process is accelerated for \textit{Paint Control}. In summary, within a case study for \ourApproach, with the same \fractionNum, smaller \subproblemSize tends to result in faster convergence than larger ones.

\begin{figure*}[!tb]
\centering
\begin{subfigure}{.19\textwidth}
\centering
\includegraphics[width=1\linewidth]{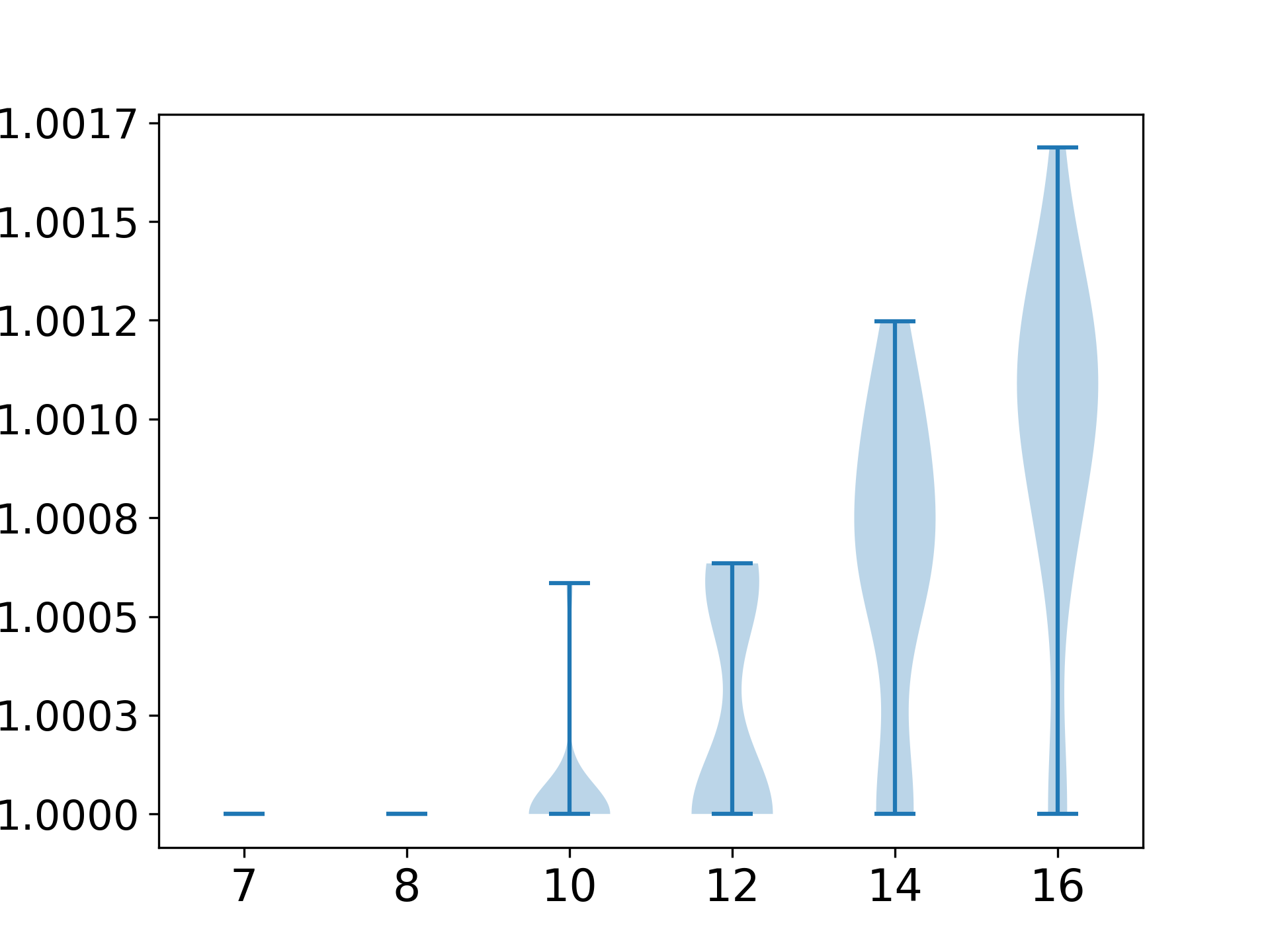}
\caption{\revision{Paint Control}}
\label{fig:finalfval_sub1}
\end{subfigure}%
\begin{subfigure}{.19\textwidth}
\centering
\includegraphics[width=1\linewidth]{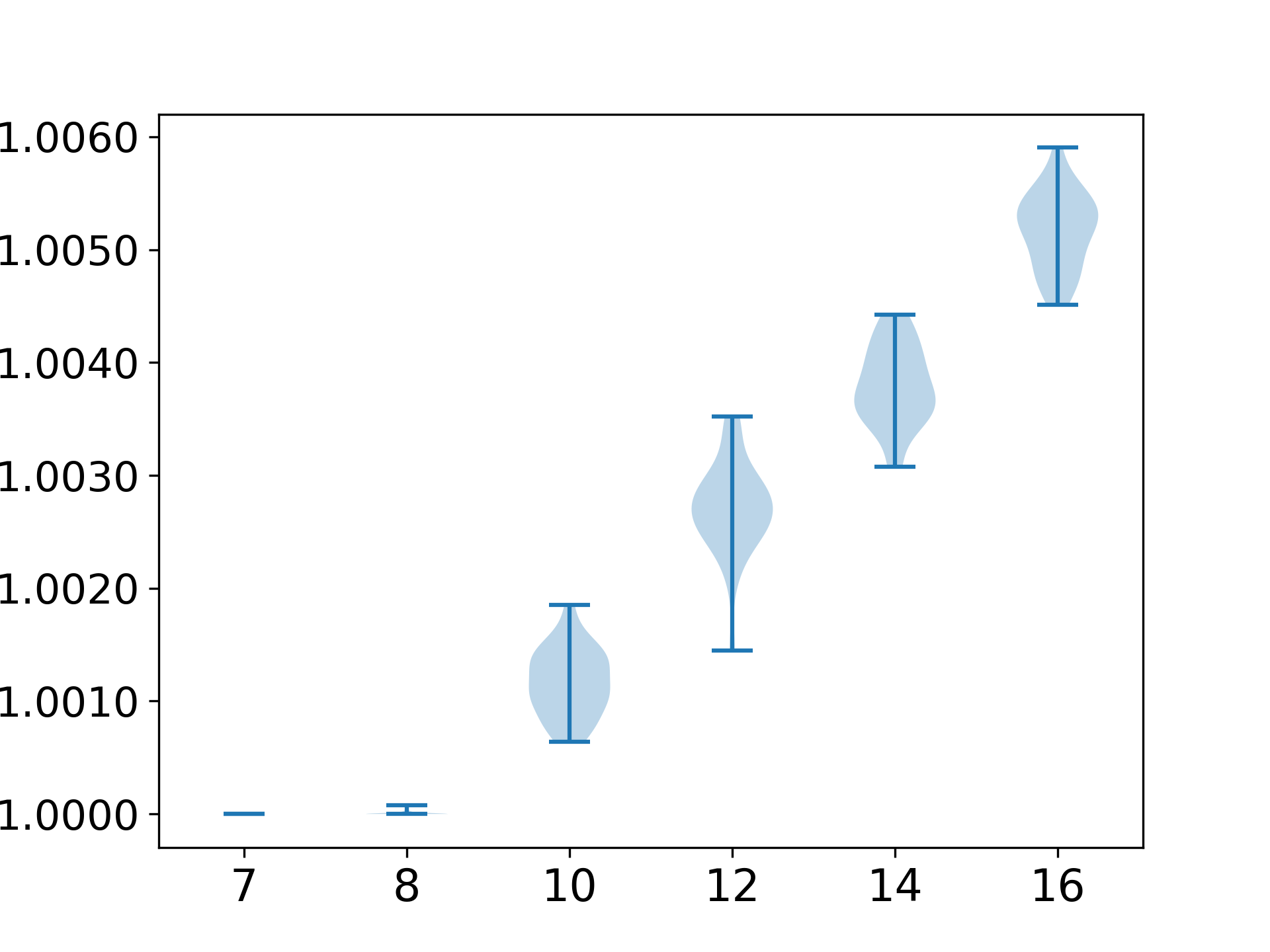}
\caption{\revision{IOF/ROL}}
\label{fig:finalfval_sub2}
\end{subfigure}
\begin{subfigure}{.19\textwidth}
\centering
\includegraphics[width=1\linewidth]{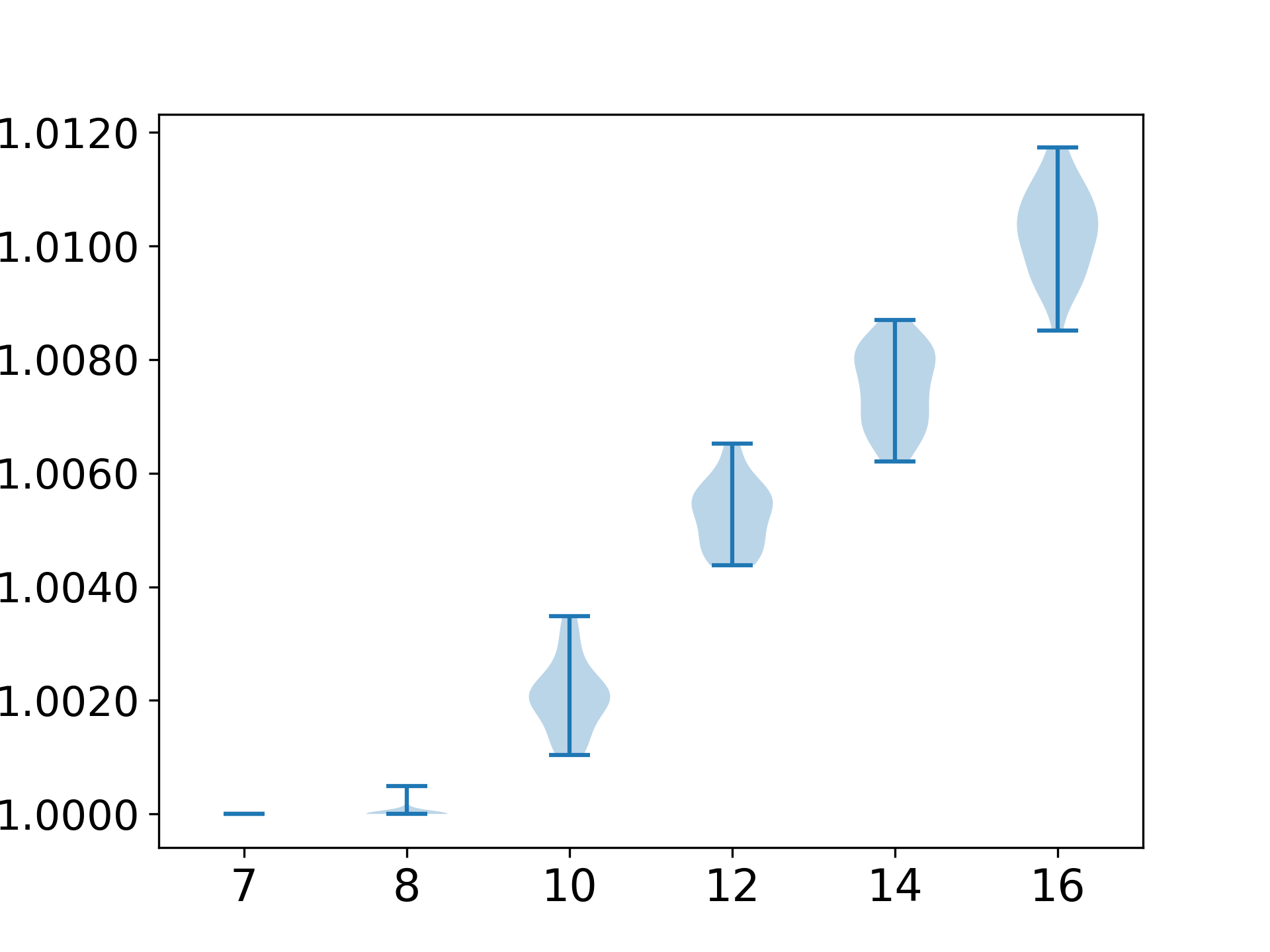}
\caption{\revision{ELEVATOR$_{o2}$}}
\label{fig:finalfval_sub3}
\end{subfigure}
\begin{subfigure}{.19\textwidth}
\centering
\includegraphics[width=1\linewidth]{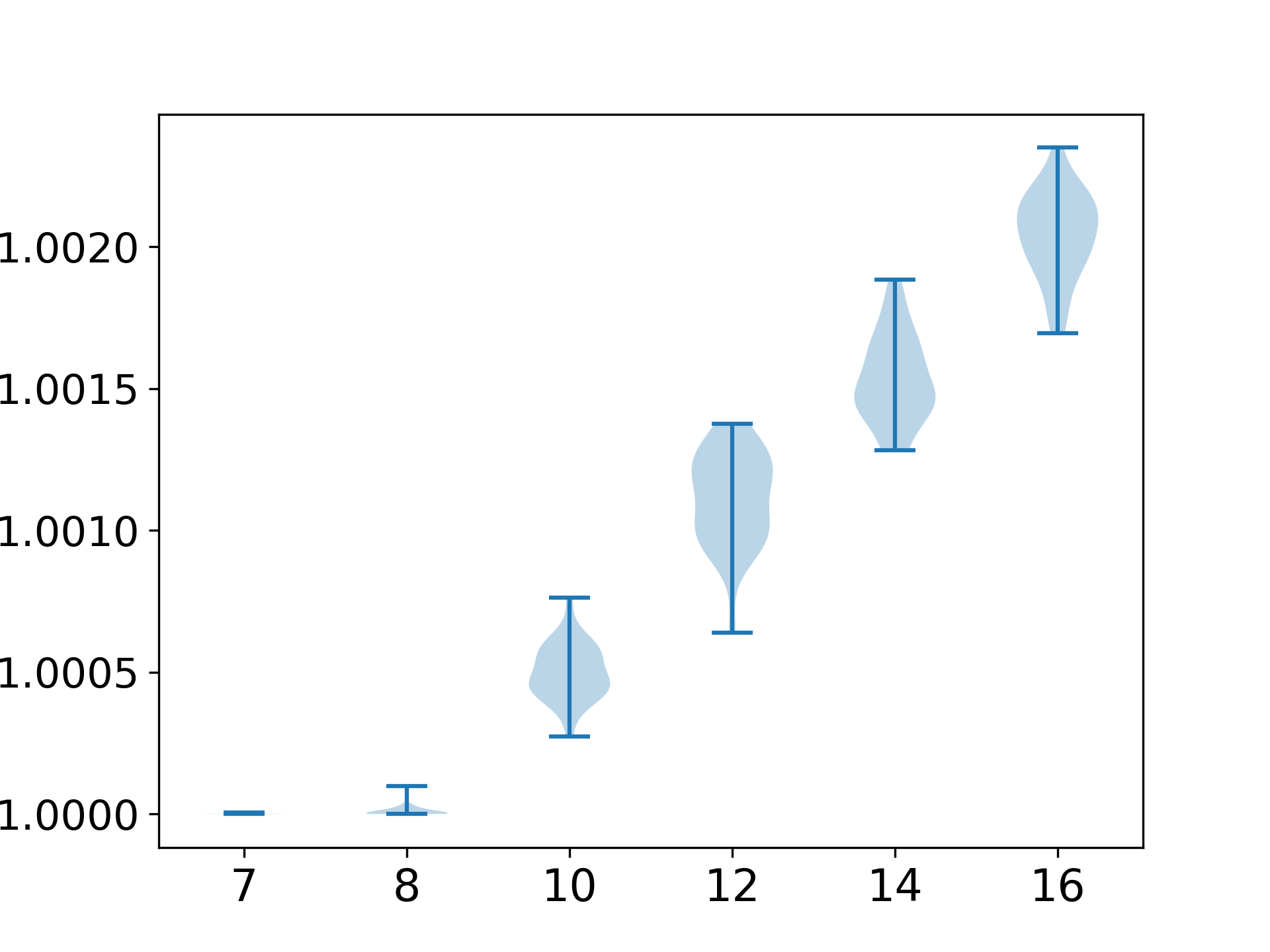}
\caption{\revision{ELEVATOR$_{o3}$}}
\label{fig:finalfval_sub4}
\end{subfigure}
\begin{subfigure}{.19\textwidth}
\centering
\includegraphics[width=1\linewidth]{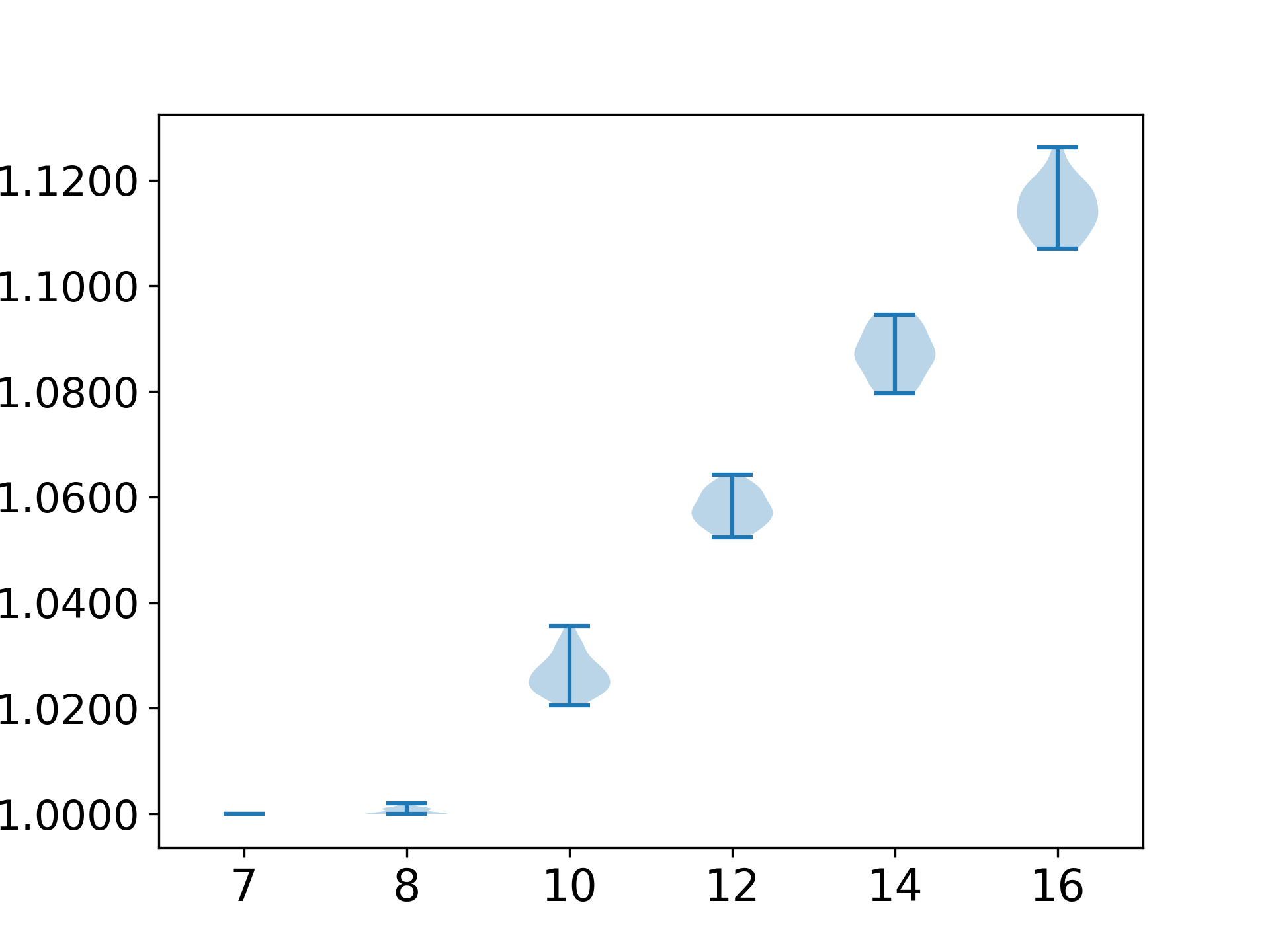}
\caption{\revision{GSDTSR}}
\label{fig:finalfval_sub5}
\end{subfigure}
\caption{\revision{Violin plots of \appro across different \subproblemSize of \ourApproach -- RQ1. The x-axis shows the sub-problem sizes (i.e., \subproblemSize of 7, 8, 10, 12, 14, and 16), whereas the y-axis shows the approximation ratio \appro}}
\label{fig:finalfval}
\end{figure*}

In Figs.~\ref{fig:finalfval_sub1}-\ref{fig:finalfval_sub5}, we show the violin plots corresponding to the \revision{30} runs of \ourApproach with each \subproblemSize for all five case studies.
Note that we pick the best \appro of each run (i.e., the lowest value) across all the iterations. From the figure, we can observe that for \textit{Paint Control}, \ourApproach with smaller \subproblemSize (i.e., 7 and 8) found the optimal solution 1.0. A similar pattern is observed for the other case studies when \ourApproach takes \subproblemSize of 7. Moreover, for \textit{Paint Control}, \ourApproach with \subproblemSize of 10, 12, 14, and 16, we can easily observe higher \appro values with a certain level of variances. For the other four case studies, \appro values obtained by \ourApproach with \subproblemSize over 8 are higher than the optimal solutions. Note that the differences among the optimal solutions obtained by \ourApproach with different \subproblemSize are not high. \revision{For instance, for IOF/ROL, \ourApproach with $\subproblemSize=16$ achieved, on average, a value of \appro 0.00517 higher than the optimal value 1.0.}

We also conduct statistical tests to compare the performance of \ourApproach, in terms of \appro, when configured with different \subproblemSize for each case study. The Kruskal-Wallis H-test results show that all p-values are smaller than 0.05 for each case study. Then, for each pair of \subproblemSize, we use the Mann-Whitney U test and Vargha and Delaney's \Atwelve effect size. Consistently, for all case studies, \ourApproach with $N=7$ statistically outperforms the other \subproblemSize (except for Paint Control for which \ourApproach with $N=7, 8$ shares the same effectiveness). Also, \ourApproach with a lower \subproblemSize is consistently more effective than a larger \subproblemSize, and the difference is statistically significant. Thus, we conclude that with lower sub-problem sizes (e.g., 7 and 8), \ourApproach tends to find more optimal solutions. 


Finally, we study the execution cost of \ourApproach with different \subproblemSize. Its total execution time consists of QAOA execution time and other computation time spent, mainly on calculating the impact order. Results are summarized in Table~\ref{table:our_approach_time}.
\begin{table*}[!tb]
\caption{Execution time of \ourApproach with different \subproblemSize -- RQ1. The total execution time (i.e., \textit{exTime} in seconds) is shown in the format of \textit{average} $\pm$ \textit{standard deviation}; \textit{QAOA\%} shows the percentage of execution time out of the total execution time spent on executing QAOA}
\label{table:our_approach_time}
\resizebox{\textwidth}{!}{
\begin{tabular}{c|c|c|c|c|c|c|c|c|c|c|c|c}
\toprule
\multicolumn{1}{c|}{\multirow{2}{*}{\centering \textbf{Case Study}}} & \multicolumn{2}{c|}{$\subproblemSize=7$} & \multicolumn{2}{c|}{$\subproblemSize=8$} & \multicolumn{2}{c|}{$\subproblemSize=10$} & \multicolumn{2}{c|}{$\subproblemSize=12$} & \multicolumn{2}{c|}{$\subproblemSize=14$} & \multicolumn{2}{c}{$\subproblemSize=16$} \\
\cline{2-13}
& \textit{exTime (s)} & QAOA\% & \textit{exTime (s)} & QAOA\% & \textit{exTime (s)} & QAOA\% & \textit{exTime (s)} & QAOA\% & \textit{exTime (s)} & QAOA\% & \textit{exTime (s)} & QAOA\% \\
\midrule
\textbf{Paint Control} & \revision{7.5$\pm$0.7} &\revision{96.2\%} &\revision{9.8$\pm$1.0} &\revision{97.2\%} &\revision{19.1$\pm$2.6} &\revision{98.3\%} &\revision{28.5$\pm$3.7} &\revision{98.7\%} &\revision{54.4$\pm$8.5} &\revision{99.2\%} &\revision{56.3$\pm$9.3} &\revision{99.2\%}\\
\textbf{IOF/ROL} & \revision{260.0$\pm$26.5} &\revision{92.5\%} &\revision{311.5$\pm$39.5} &\revision{94.4\%} &\revision{463.4$\pm$67.1} &\revision{96.8\%} &\revision{668.3$\pm$118.8} &\revision{97.8\%} &\revision{1078.5$\pm$170.6} &\revision{98.8\%} &\revision{1160.4$\pm$272.1} &\revision{98.7\%} \\
\textbf{ELEVATOR$_{o2}$} & \revision{264.9$\pm$21.3} &\revision{93.0\%} &\revision{343.9$\pm$108.3} &\revision{94.8\%} &\revision{531.8$\pm$333.8} &\revision{97.0\%} &\revision{733.3$\pm$350.3} &\revision{97.9\%} &\revision{1274.0$\pm$645.1} &\revision{98.8\%} &\revision{15309.4$\pm$19040.3} &\revision{99.9\%} \\
\textbf{ELEVATOR$_{o3}$} & \revision{302.3$\pm$34.1} &\revision{92.4\%} &\revision{349.0$\pm$48.0} &\revision{94.3\%} &\revision{473.7$\pm$74.9} &\revision{96.8\%} &\revision{681.1$\pm$136.4} &\revision{97.8\%} &\revision{1204.5$\pm$306.7} &\revision{98.7\%} &\revision{1594.9$\pm$2823.1} &\revision{98.7\%} \\
\textbf{GSDTSR} & \revision{540.0$\pm$39.2} &\revision{83.1\%} &\revision{673.8$\pm$66.5} &\revision{87.5\%} &\revision{1066.8$\pm$199.2} &\revision{92.9\%} &\revision{1568.4$\pm$861.9} &\revision{94.7\%} &\revision{1580.5$\pm$427.7} &\revision{96.0\%} &\revision{2055.3$\pm$2741.2} &\revision{95.8\%} \\
\bottomrule
\end{tabular}
}
\end{table*}
There are two key observations. First, for all the case studies, \ourApproach with higher \subproblemSize incurs higher execution cost, even though \ourApproach with smaller \subproblemSize needs more QAOA runs. This is because executing large problems with QAOA on the simulator takes much longer than solving smaller problems. Second, QAOA takes the most execution time (e.g., 96.2\% for Paint Control when \subproblemSize=7), which also increases when \subproblemSize increases. This is because the other portion of the execution time (mainly about applying the IGDec strategy) does not change much when \subproblemSize changes. It is also important to mention that the current time performance of \ourApproach is based on executing QAOA on the simulator. When they are executed on real quantum computers, it is expected to be significantly more efficient. 

\begin{tcolorbox}[size=title, colframe=green!10, width=1\linewidth, colback=green!10,breakable]
\textbf{Conclusions for RQ1:} 
There is no significant difference between different numbers of layers of QAOA in terms of the performance of \ourApproach for test case optimization. Thus, we suggest using \textit{p=1}, which incurs minimum cost. \ourApproach with smaller \subproblemSize tends to achieve better effectiveness. 
Thus, we suggest using lower \subproblemSize for test optimization with \ourApproach.
\end{tcolorbox}


\subsection{RQ2 -- Comparing \ourApproach with Baseline}
RQ2 explores the benefits of employing the IGDec strategy of \ourApproach. To this end, we choose the best parameter settings of \ourApproach, i.e., $p=1$ and $N=7$. We use the results of \revision{30 runs} of \ourApproach and \baselineqaoa and compare them in terms of \appro.

\begin{figure*}[!tb]
\centering
\begin{subfigure}{.19\textwidth}
\centering
\includegraphics[width=1\linewidth]{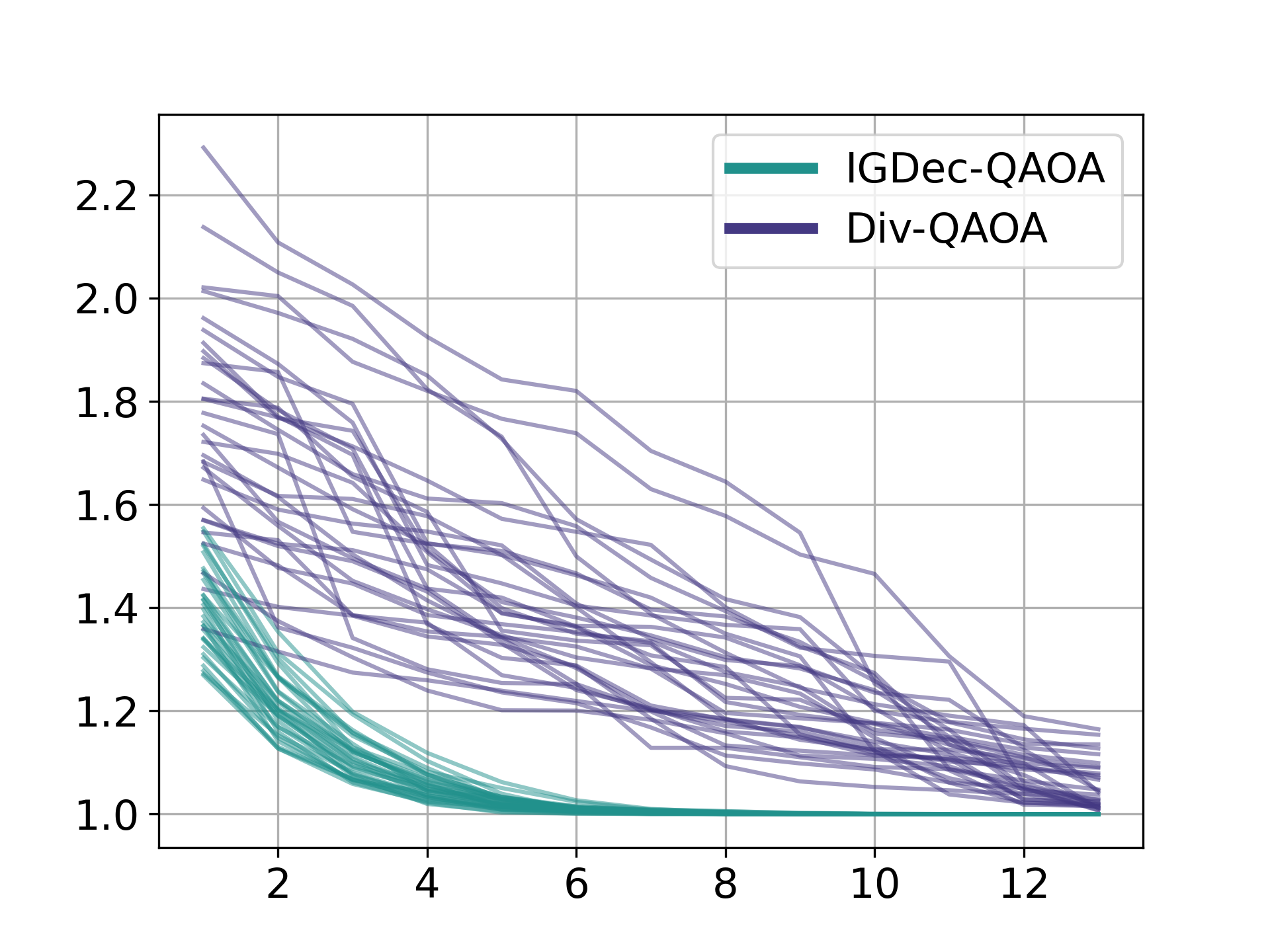}
\caption{\revision{Paint Control}}
\label{fig:basaeline_trendsub1}
\end{subfigure}%
\begin{subfigure}{.19\textwidth}
\centering
\includegraphics[width=1\linewidth]{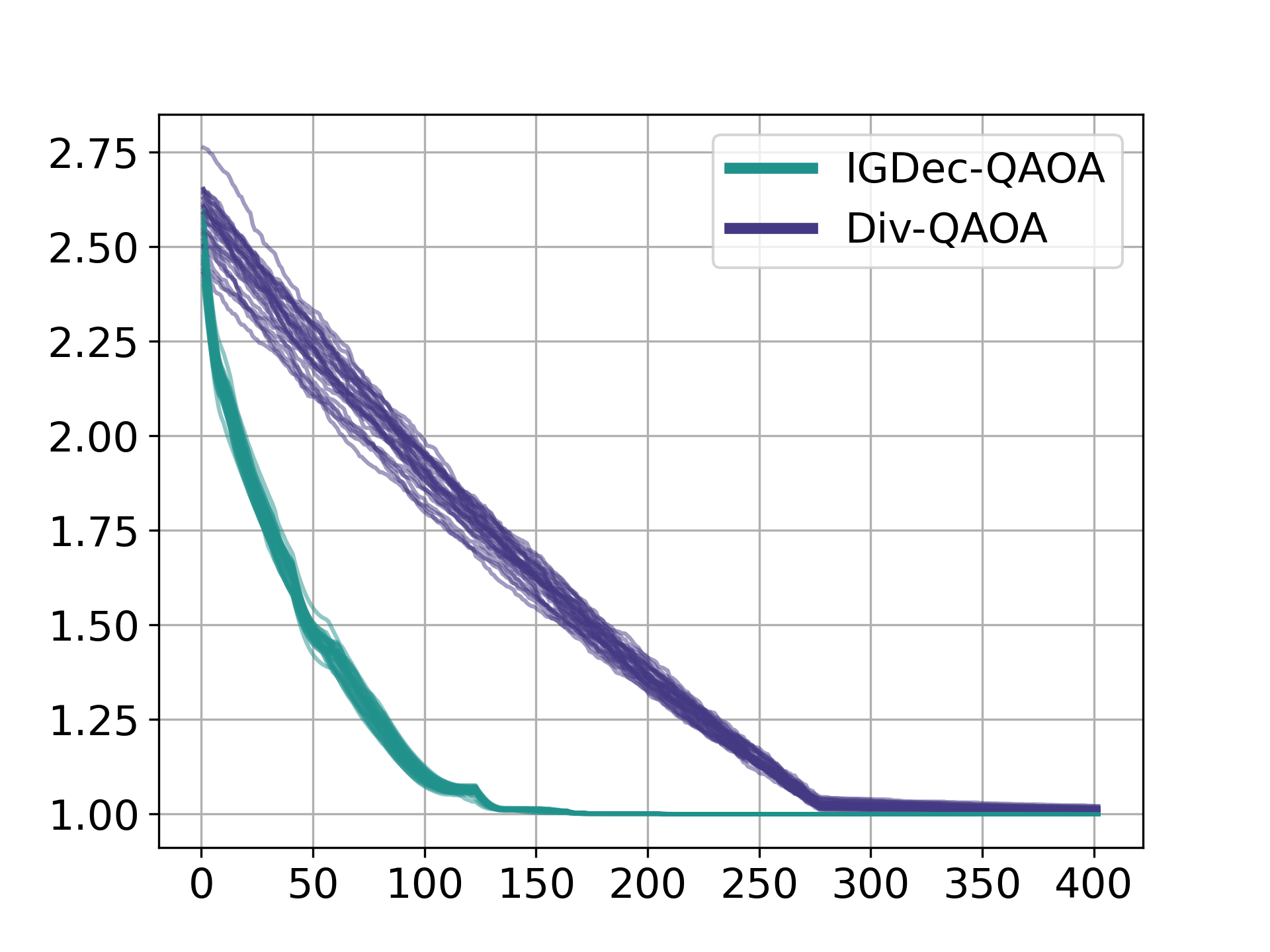}
\caption{\revision{IOF/ROL}}
\label{fig:basaeline_trendsub2}
\end{subfigure}
\begin{subfigure}{.19\textwidth}
\centering
\includegraphics[width=1\linewidth]{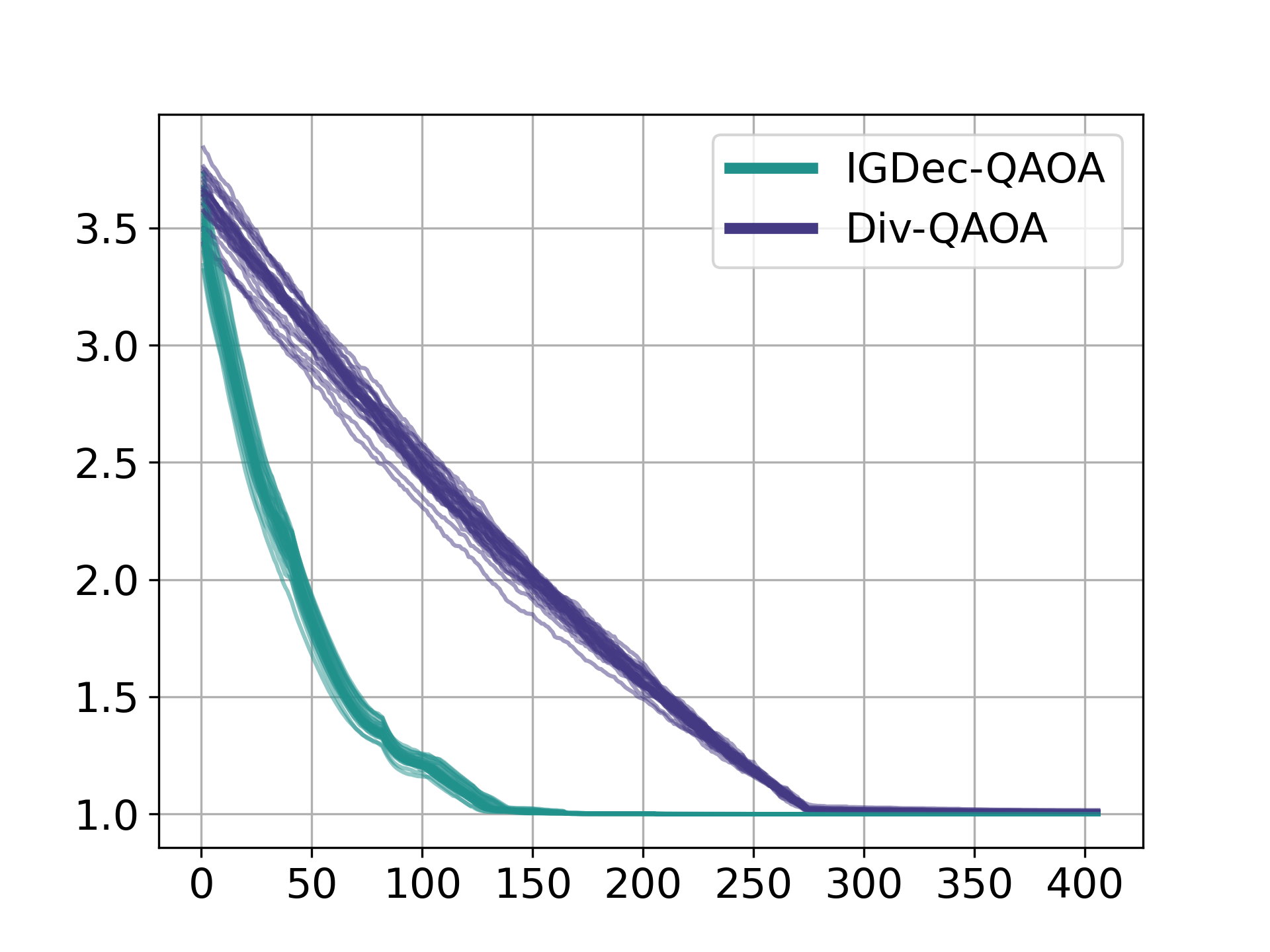}
\caption{\revision{ELEVATOR$_{o2}$}}
\label{fig:basaeline_trendsub3}
\end{subfigure}
\begin{subfigure}{.19\textwidth}
\centering
\includegraphics[width=1\linewidth]{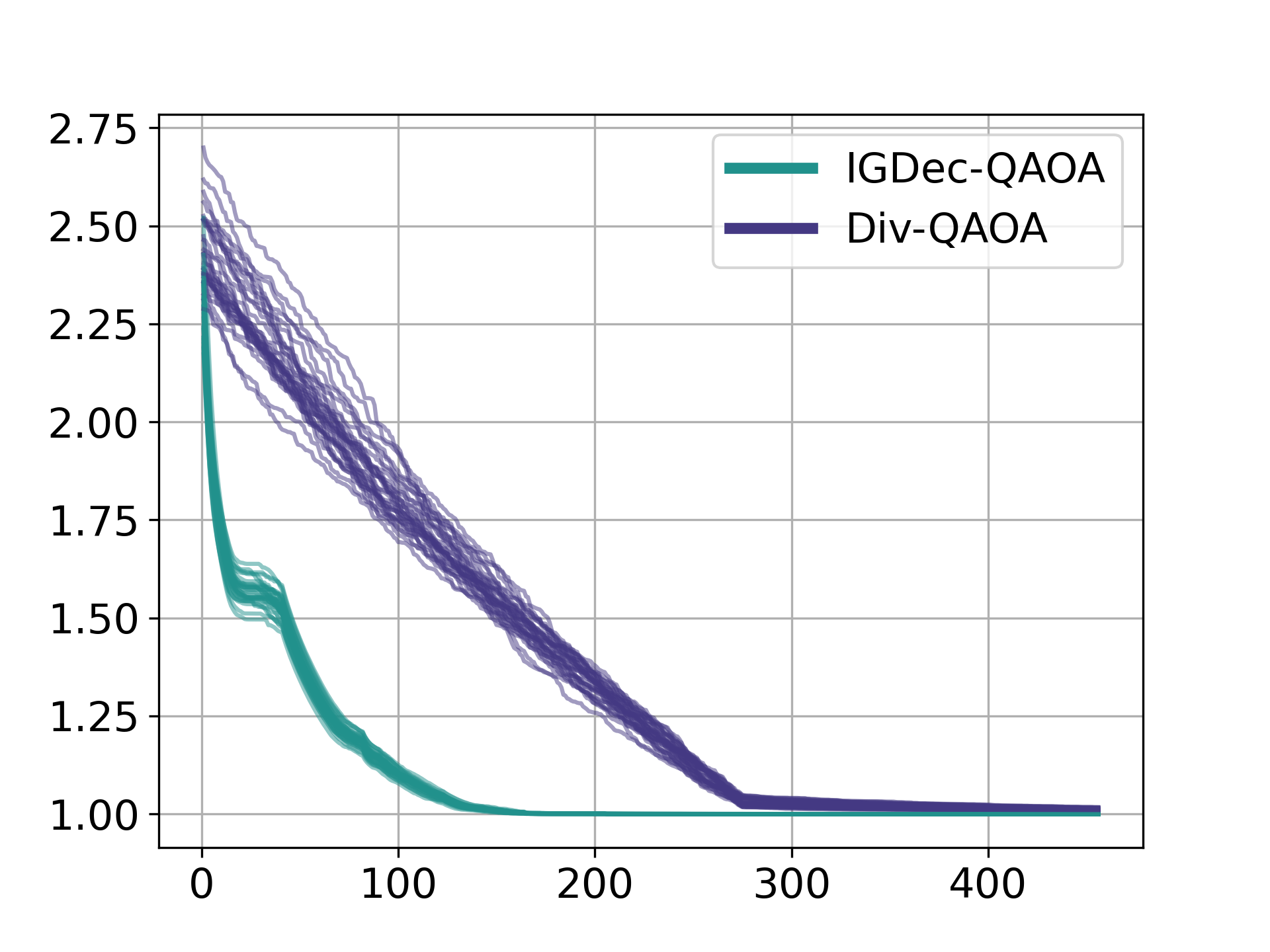}
\caption{\revision{ELEVATOR$_{o3}$}}
\label{fig:basaeline_trendsub4}
\end{subfigure}
\begin{subfigure}{.19\textwidth}
\centering
\includegraphics[width=1\linewidth]{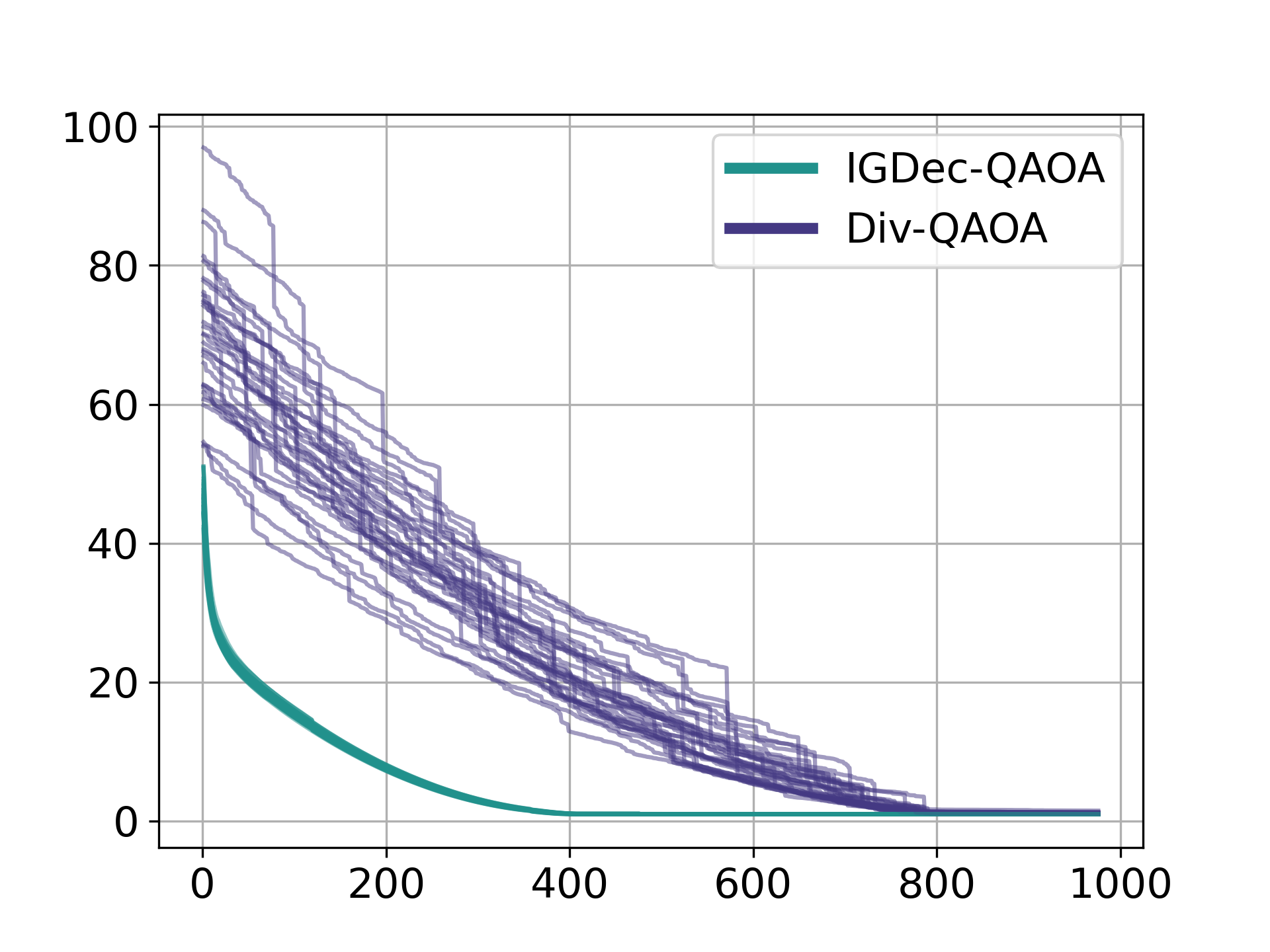}
\caption{\revision{GSDTSR}}
\label{fig:basaeline_trend_sub5}
\end{subfigure}
\caption{\revision{Approximation ratio (\appro) trends of \ourApproach and \baselineqaoa along with the QAOA executions (\numEval) for five case studies -- RQ2}}
\label{fig:basaeline_trend}
\end{figure*}

Fig.~\ref{fig:basaeline_trend} (a)-(e) compare the two approaches across all five case studies.
Note that both approaches use the same number of QAOA executions (the total number of sub-problems being solved by QAOA in all iterations, i.e., \numEval) to ensure a fair comparison. Regarding the overall trends, one can observe that the \appro values of the two approaches all decrease rapidly during the initial half of the QAOA executions and then stabilize. However, \ourApproach exhibits a faster decrease for all case studies because of its impact ordering strategy; \ourApproach orders test cases in each iteration, contributing to the reduction of fitness values. Conversely, in \baselineqaoa, although each variable (i.e., test case) is optimized at least once into the \activeSet, the sequence of optimization is determined randomly, lacking an efficient order for optimization. Thus, the absence of the IGDec strategy results in a slower decline of fitness values and, consequently, of the \appro values as well. 

\begin{figure*}[!tb]
\centering
\begin{subfigure}{.19\textwidth}
\centering
\includegraphics[width=1\linewidth]{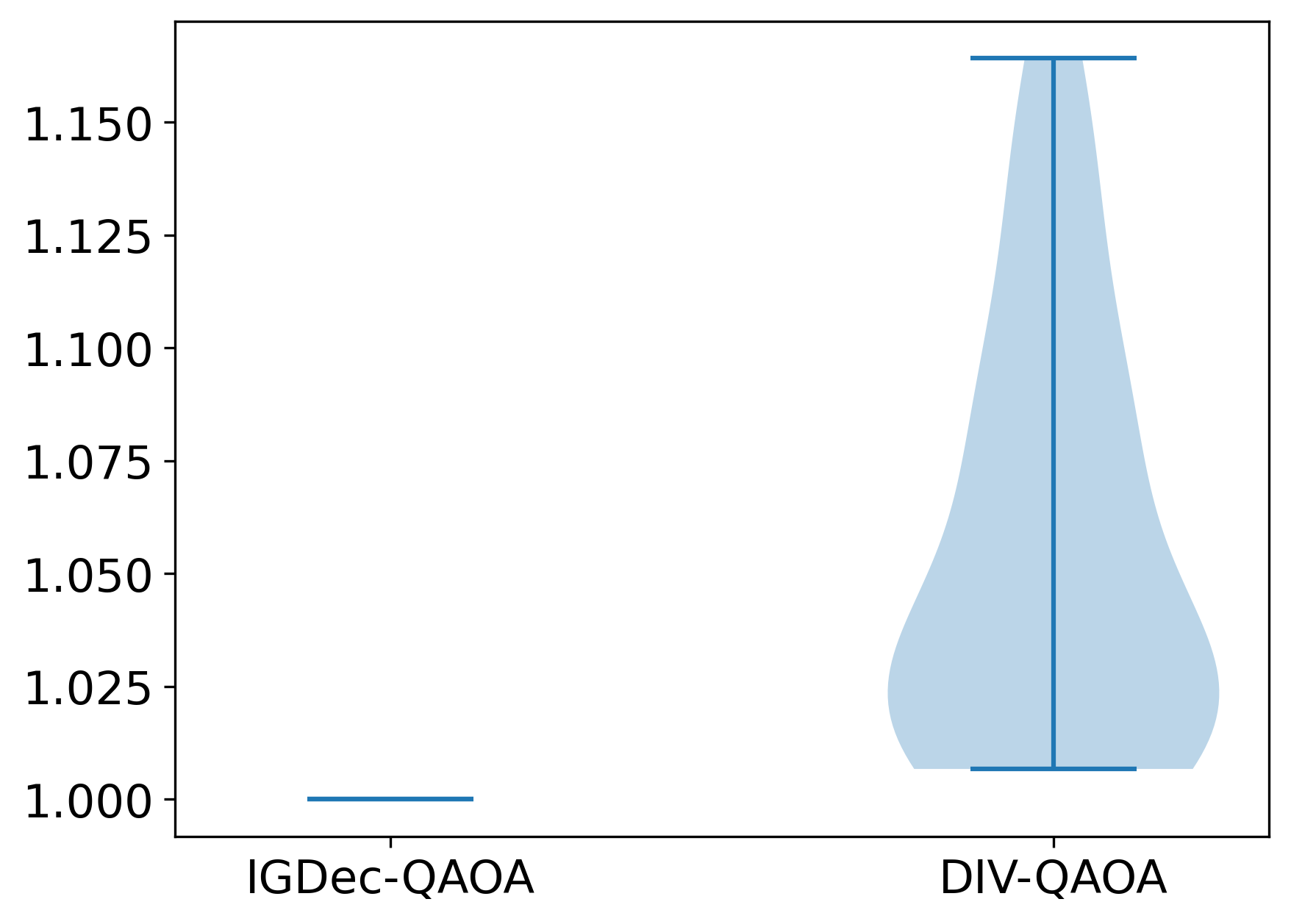}
\caption{\revision{Paint Control}}
\label{fig:baseline_box_sub1}
\end{subfigure}%
\begin{subfigure}{.19\textwidth}
\centering
\includegraphics[width=1\linewidth]{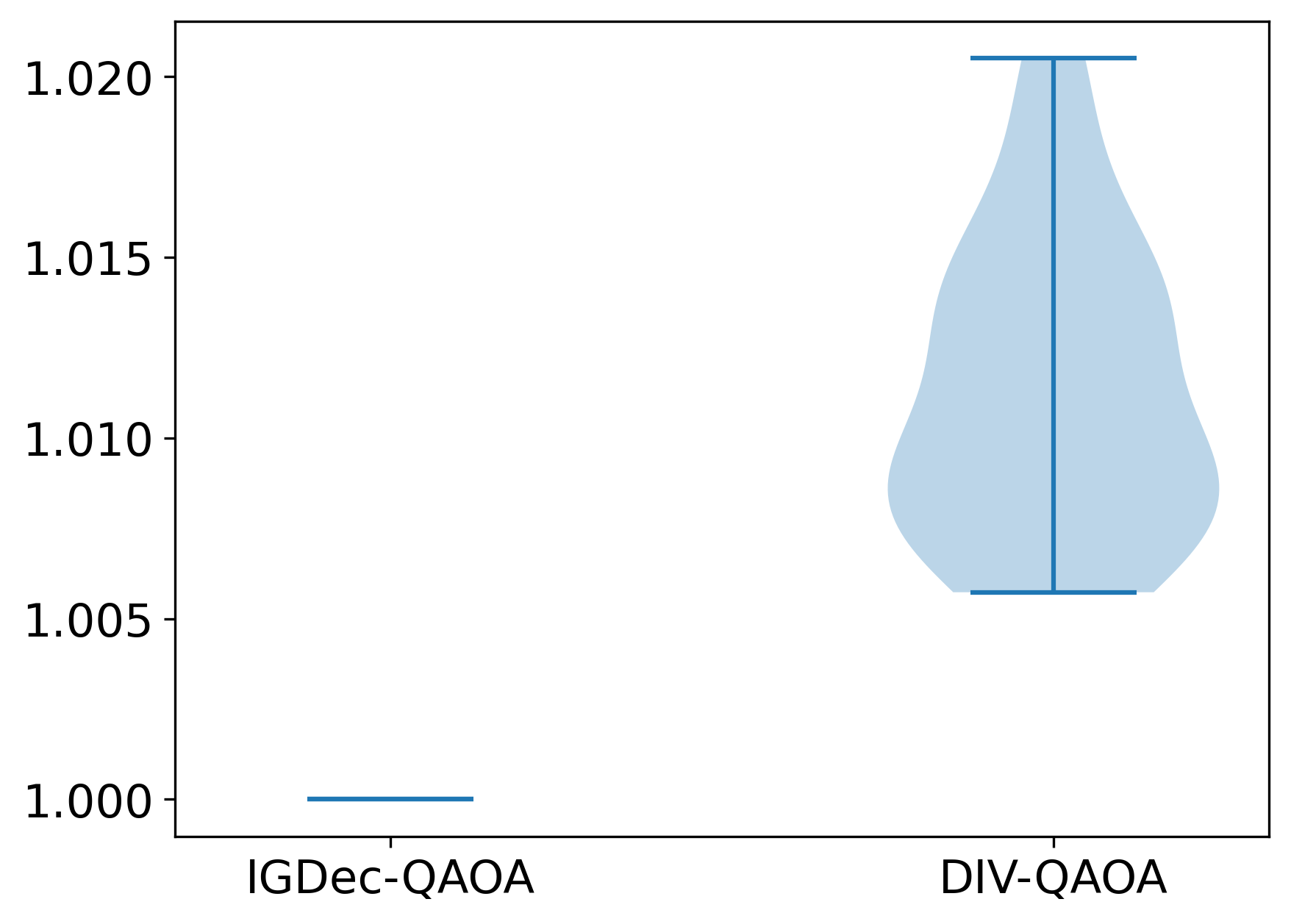}
\caption{\revision{IOF/ROL}}
\label{fig:baseline_box_sub2}
\end{subfigure}
\begin{subfigure}{.19\textwidth}
\centering
\includegraphics[width=1\linewidth]{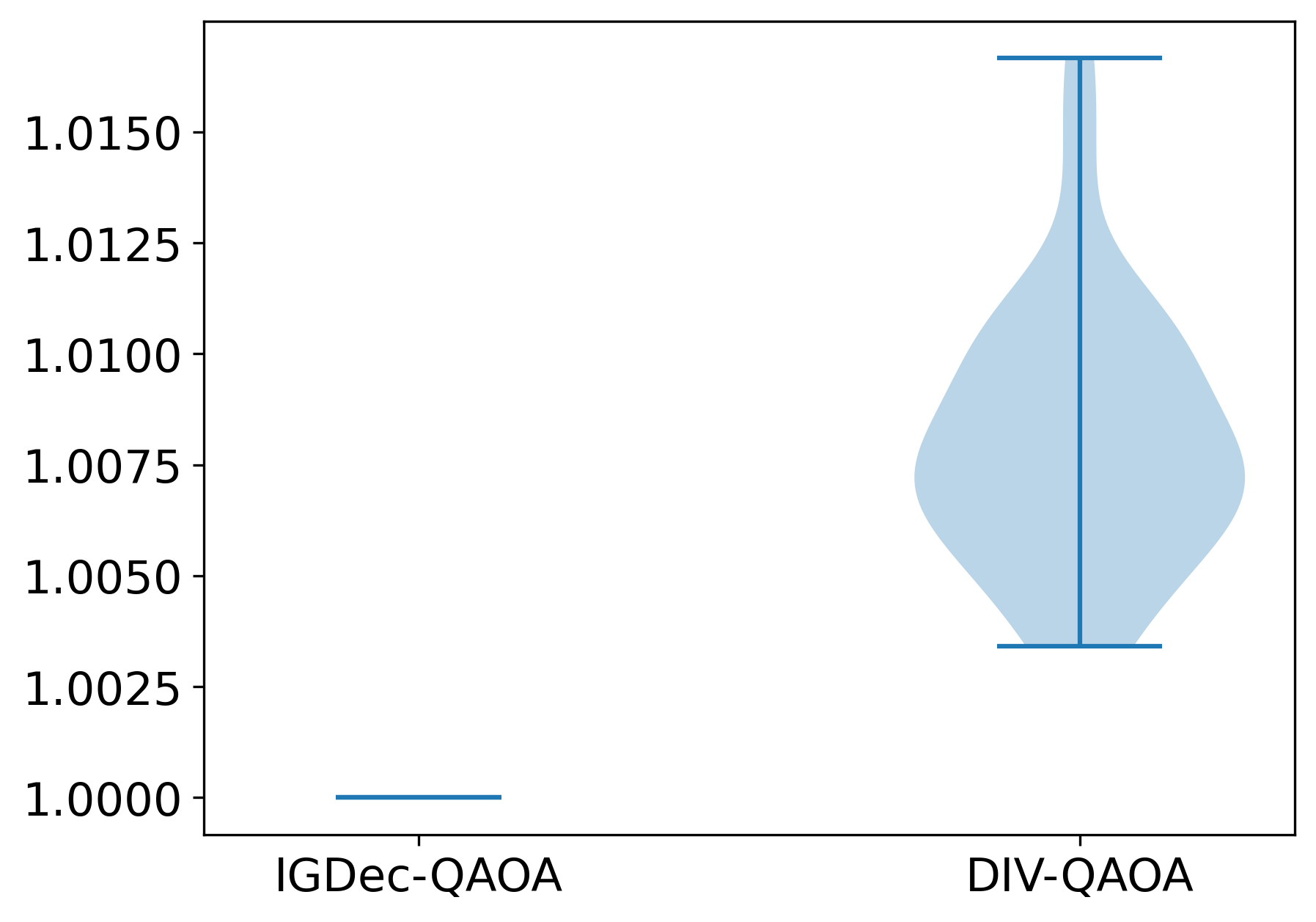}
\caption{\revision{ELEVATOR$_{o2}$}}
\label{fig:baseline_box_sub3}
\end{subfigure}
\begin{subfigure}{.19\textwidth}
\centering
\includegraphics[width=1\linewidth]{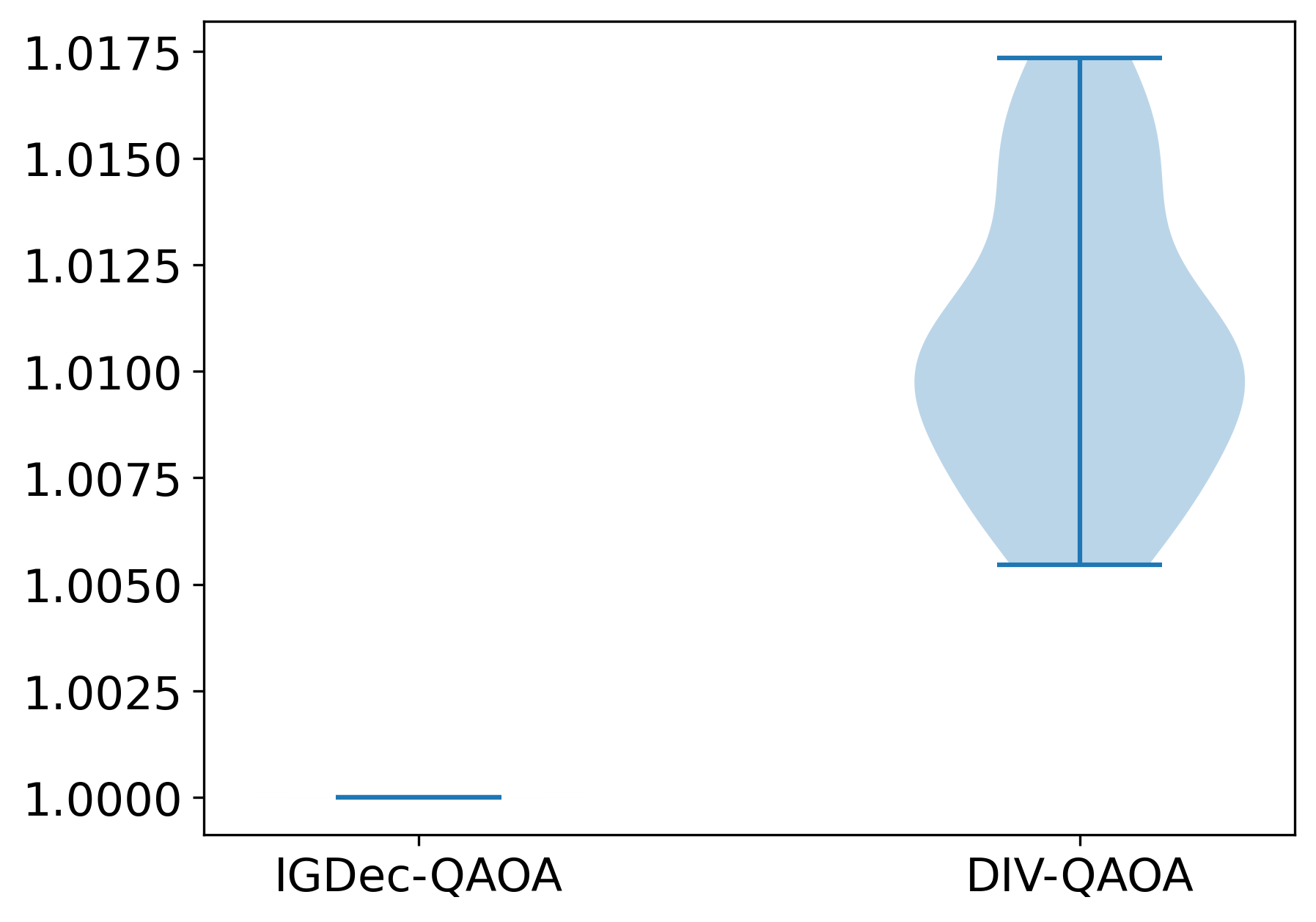}
\caption{\revision{ELEVATOR$_{o3}$}}
\label{fig:baseline_box_sub4}
\end{subfigure}
\begin{subfigure}{.19\textwidth}
\centering
\includegraphics[width=1\linewidth]{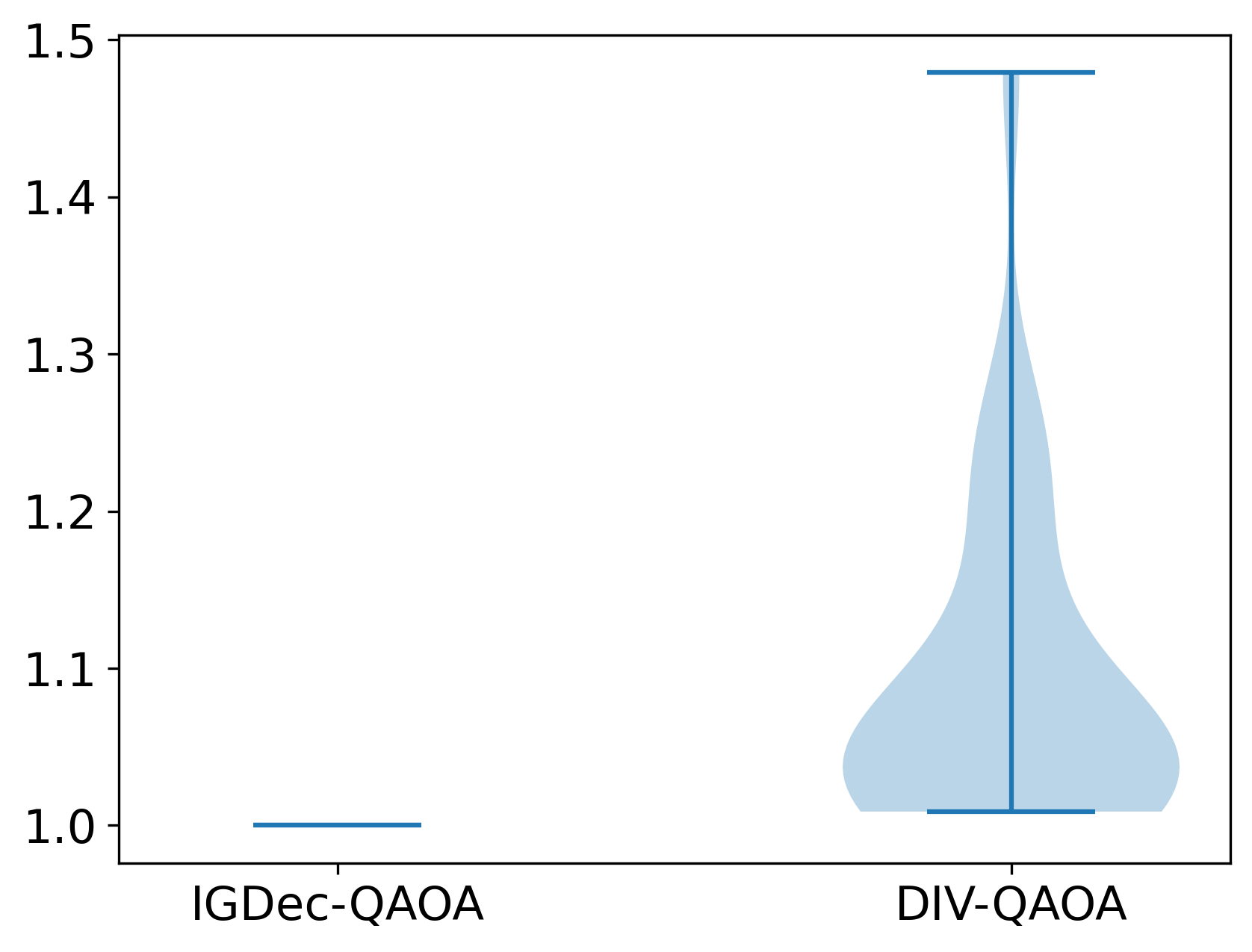}
\caption{\revision{GSDTSR}}
\label{fig:baseline_box_sub5}
\end{subfigure}
\caption{\revision{Violin plots of the final approximation ratio (\appro) of \ourApproach and \baselineqaoa for five case studies -- RQ2}}
\label{fig:baseline_box}
\end{figure*}

Fig.~\ref{fig:baseline_box} shows that \ourApproach always manages to reach the optimal value 1.0 throughout all \revision{30 runs} for all five case studies.
In comparison, \baselineqaoa consistently performs worse than \ourApproach. In addition, results of the statistical test we conducted tell that \ourApproach significantly outperforms \baselineqaoa for all case studies. These results indicate that the random selection of test cases for each sub-problem in each execution is ineffective for optimization, thereby requiring the IGDec strategy in \ourApproach. 

Additionally, as depicted in Sect.~\ref{subsec:RQ1results}, the time consumed by \ourApproach, excluding QAOA execution, is relatively insignificant compared to the time required for QAOA execution. In summary, \ourApproach outperforms \baselineqaoa both in terms of effectiveness and time cost.
\begin{tcolorbox}[size=title, colframe=green!10, width=1\linewidth, colback=green!10,breakable]
\textbf{Conclusions for RQ2:} 
\ourApproach significantly outperforms \baselineqaoa in terms of finding approximate optimal solutions, thereby suggesting that combining the impact ordering strategy with QAOA is promising in solving test optimization problems.
\end{tcolorbox}

\subsection{RQ3 -- Comparing \ourApproach with classical Algorithms}\label{subsec:rq4}
In RQ3, we compare \ourApproach (configured with $p=1$ and $N=7$) with RS and GA. For each problem, we run RS with the average number of QAOA executions \numEval of \ourApproach as the iteration count, and we use the lowest fitness values obtained as the solution to calculate \appro. Results are shown in Table~\ref{tab:ga_comparison}.

RS fails to achieve the optimal \appro value 1.0 in all runs for all case studies. Especially for the largest case study \textit{GSDTSR}, the \appro value achieved by RS is as high as 44.881. The results of the Mann-Whitney U test revealed all p-values less than 0.05, signifying that RS is significantly worse than \ourApproach for all case studies. Moreover, \Atwelve values are less than 0.01; indicating strong differences. This result suggests that the TCO problems are difficult to solve, justifying the need for optimization algorithms.
\begin{table*}[!tb]
\caption{Results of the performance of \ourApproach and GA -- RQ3. Column $\textbf{avg}(\appro)-1.0$ show the difference between the average \appro across \revision{30 runs} and the optimal (i.e., 1.0); \textbf{std$(\appro)$} show standard deviations of \revision{30 runs}; \pop refers to the optimal population size of GA; \textbf{\numEval} refers to the average values; \textbf{fail} lists the number of runs that fail to achieve the optimal \appro 1.0; \textbf{Result} presents comparison between \ourApproach and GA.}
\resizebox{\textwidth}{!}{
\begin{tabular}{c|c|c|c|c|c|c|c|c|c|c|c|c|c|c} 
\toprule
\multirow{2}{*}{\textbf{Case}} & \multicolumn{4}{c|}{\textbf{\ourApproach}} & \multicolumn{4}{c|}{\textbf{RS}} & \multicolumn{5}{c|}{\textbf{GA}} & \multirow{2}{*}{\textbf{Result}} \\
\cline{2-5}\cline{6-9}\cline{10-14}
& \textbf{avg$(\appro)-1.0$} & \textbf{std$(\appro)$} & \textbf{fail} & \numEval & \textbf{avg$(\appro)-1.0$} & \textbf{std$(\appro)$} & \textbf{fail} & \numEval & \textbf{avg$(\appro)-1.0$} & \textbf{std$(\appro)$} & \textbf{fail} & \pop & \numEval & \\ 
\midrule
Paint Control & 0 & 0 & 0 & \revision{11.6} & \revision{0.473} & \revision{0.079} & \revision{30} & \revision{12} & 0 & 0 & 0 & \revision{10} & \revision{2,530} & $\equiv$ \\
IOF/ROL & 0 & 0 & 0 & \revision{400.4} & \revision{1.387} & \revision{0.024} & \revision{30} & \revision{401} & \revision{\num{6.223e-07}} & \revision{\num{9.495e-07}} & \revision{16} & \revision{30} & \revision{28,670} & \checkmark \\
ELEVATOR$_{o2}$ & 0 & 0 & 0 & \revision{412.7} & \revision{2.348} & \revision{0.030} & \revision{30} & \revision{413} & \revision{\num{2.067e-06}} & \revision{\num{3.860e-06}} & \revision{19} & \revision{80} & \revision{135,600} & \checkmark \\
ELEVATOR$_{o3}$ & \revision{\num{1.355e-06}} & \revision{\num{1.916e-06}} & \revision{10} & \revision{463.3} & \revision{1.187} & \revision{0.023} & \revision{30} & \revision{464} & \revision{\num{1.608e-06}} & \revision{\num{2.439e-06}} & \revision{10} & \revision{90} & \revision{145,962} & $\equiv$ \\
GSDTSR & 0 & 0 & 0 & \revision{959.9} & \revision{44.881} & \revision{1.472} & \revision{30} & \revision{960} & 0 & 0 & 0 & \revision{50} & \revision{17,744} & $\equiv$ \\
\bottomrule
\end{tabular}
}
\label{tab:ga_comparison}
\begin{flushleft}
In the \textbf{Result} column: $\equiv$ denotes no statistically significant difference between \ourApproach and GA, while \checkmark denotes \ourApproach performs significantly better than GA.
\end{flushleft}
\end{table*}

As described in Sect.~\ref{sec:baseline}, we run GA with various population sizes \pop, and for each experiment, we run 400,000 evaluations, sufficient to observe when GA converges. For each dataset, we calculate the average lowest fitness value among \revision{30 runs} achieved by each \pop and also report the average number of evaluations \numEval GA takes to obtain the lowest fitness values in that run. We select the optimal \pop that achieved the lowest fitness value with the smallest \numEval for each dataset. The information of the optimal \pop values and their corresponding \numEval are reported in Table~\ref{tab:ga_comparison}. The average \appro and standard deviation values of those \revision{30 runs} are also reported, together with the number of runs that failed to obtain the optimal \appro value (i.e., 1.0). Besides, we conduct statistical tests to compare \appro of \revision{30 runs} achieved by our approach and GA with optimal \pop for each dataset and provide the results in Table~\ref{tab:ga_comparison}.

As shown in Table~\ref{tab:ga_comparison}, for IOF/ROL and ELEVATOR$_{o2}$, \ourApproach significantly outperforms GA and achieves 1.0 consistently for all \revision{30} runs, which is, however, not the case for GA. When looking at \numEval, those runs in which GA fails to obtain \appro 1.0, achieve the lowest fitness values with a relatively small \numEval (e.g., \revision{28,670} for IOF/ROL) as compared to 400,000 in total. Although the discrepancy between the achieved \appro and the optimal 1.0 is minor, with both the difference and standard deviation values in the magnitude of \num{e-7} and \num{e-6}, GA is trapped in local optima for hundreds of thousands of evaluations without further optimization. This slight deviation from the optimal \appro is probably due to the different selection of a small number of test cases for the final solution. \ourApproach, however, does not have this limitation. It is probably because, in the IGDec strategy, each time a current solution is updated by a sub-solution, it will jump out of a local optimum. Also, the mixing Hamiltonian part of the QAOA circuit helps explore the search space.

In 4 out of 5 case studies, \ourApproach manages to achieve the optimal \appro value 1.0 consistently (i.e., Paint Control, IOF/ROL, ELEVATOR$_{o2}$, and GSDTSR) while GA achieves 1.0 for all \revision{30} runs only in Paint Control and GSDTSR. Statistically, the effectiveness of the two approaches is equivalent in Paint Control and GSDTSR. In Paint Control, as the dataset size is relatively small, both approaches can reach 1.0 in all \revision{30} runs. In GSDTSR, although the number of test cases is large (i.e., 5555), the ``failure rate'' values of 94.8\% test cases (see Sect.~\ref{subsec:experimentsetup}) are 0\%. Thus, since one objective is to maximize the ``failure rate'', the problem will be simple for both \ourApproach and GA to solve after they filter those test cases during the optimization process within the early stages of generations; hence, both consistently achieve the optimal \appro, i.e., 1.0.

For ELEVATOR$_{o3}$, both approaches cannot achieve \appro 1.0 in all \revision{30} runs with a slight difference, which is probably because the search problem is complex, and both approaches are stuck in local optima in some runs. 
\revision{The number of failing runs in both approaches is 10 of 30 runs.}
Also, according to the statistical test, there is no significant difference between the two approaches.

\begin{tcolorbox}[size=title, colframe=green!10, width=1\linewidth, colback=green!10,breakable]
\textbf{Conclusions for RQ3:} 
RS fails to achieve the optimal \appro value for all cases, suggesting the need for search algorithms to solve TCO problems. \ourApproach can outperform or perform comparably with GA, and it is less prone to be trapped in local optima than GA. 
\end{tcolorbox}

\subsection{RQ4 -- Performance of \ourApproach on Noisy Quantum Computer}\label{subsec:discussion}
\revision{\noindent\textbf{Performance of \ourApproach on noisy simulator:}
We ran \ourApproach with the best setting (i.e., $p=1$ and $\totalVariable=7$) on a noisy simulator to analyze the results. We implement the QAOA algorithm on a noisy quantum simulator based on the quantum computer ``ibm\_brisbane'', which closely mimics gates that will be executed on a real device and noise properties on the quantum computer. The experiment is repeated 30 times. Table~\ref{tab:noise_comparison} shows the performance of \ourApproach on the noisy simulator.
\begin{table}[!tb]
\caption{\revision{RQ4 -- Results of the performance of \ourApproach on noisy simulator.}}
\resizebox{0.5\textwidth}{!}{
\begin{tabular}{c|c|c|c|c|c} 
\toprule
\multirow{2}{*}{\textbf{Case}} & \multicolumn{4}{c|}{\revision{\textbf{\ourApproach} with noise}} & \multirow{2}{*}{\textbf{Result}} \\
\cline{2-5}
& \textbf{avg$(\appro)-1.0$} & \textbf{std$(\appro)$} & \textbf{fail} & \numEval & \\
\midrule
Paint Control & \revision{0} & \revision{0} & \revision{0} & \revision{12.1} & \revision{$\equiv$} \\
IOF/ROL & \revision{0} & \revision{0} & \revision{0} & \revision{397.7} & \revision{\checkmark} \\
ELEVATOR$_{o2}$ & \revision{0} & \revision{0} & \revision{0} & \revision{404.5} & \revision{\checkmark} \\
ELEVATOR$_{o3}$ & \revision{\num{1.6256e-06}} & \revision{\num{1.991e-06}} & \revision{12} & \revision{451.0} & \revision{$\equiv$} \\
GSDTSR &\revision{0} & \revision{0} & \revision{0} & \revision{971.8} & \revision{$\equiv$} \\
\bottomrule
\end{tabular}
}
\label{tab:noise_comparison}
\begin{flushleft}
\revision{In the \textbf{Result} column: $\equiv$ denotes no statistically significant difference between \ourApproach and GA, while \checkmark denotes \ourApproach performs significantly better than GA.}
\end{flushleft}
\end{table}
All case studies except \textit{Elevator}$_{o3}$ achieve the optimal solutions in all 30 runs, which is similar to the performance of \ourApproach on the ideal simulator. We conduct statistical tests on the \appro of \ourApproach and GA: results show that we achieve performance on the noisy simulator that is equivalent to and in two cases better than, GA.}

\revision{We show the trends of \appro for all 30 runs on both ideal and noisy simulators in Fig.~\ref{fig:noise_trend} along the increasing iterations.
\begin{figure*}[!tb]
\centering
\begin{subfigure}{.19\textwidth}
\centering
\includegraphics[width=1\linewidth]{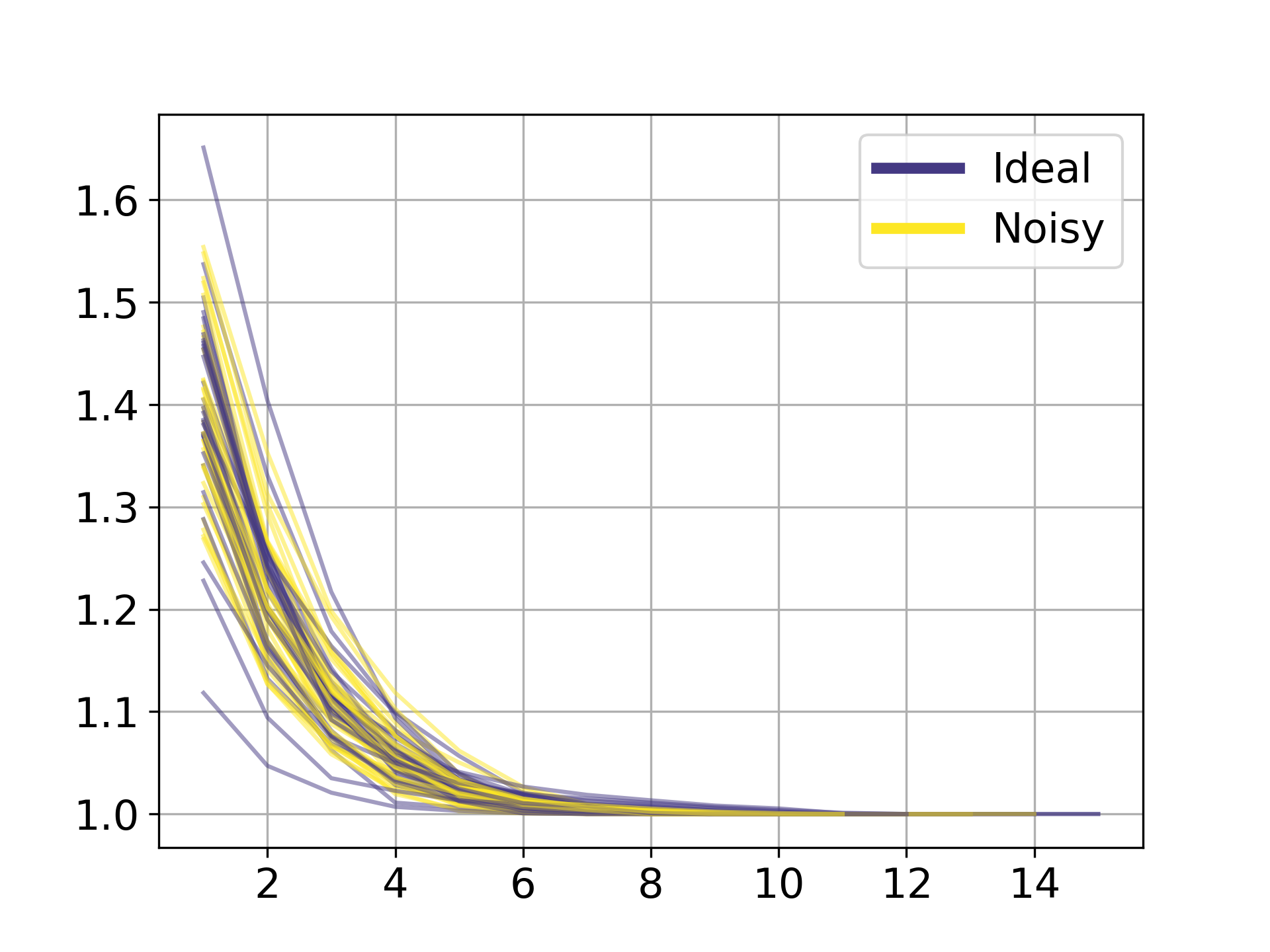}
\caption{\revision{Paint Control}}
\label{fig:noise_trend_sub1}
\end{subfigure}%
\begin{subfigure}{.19\textwidth}
\centering
\includegraphics[width=1\linewidth]{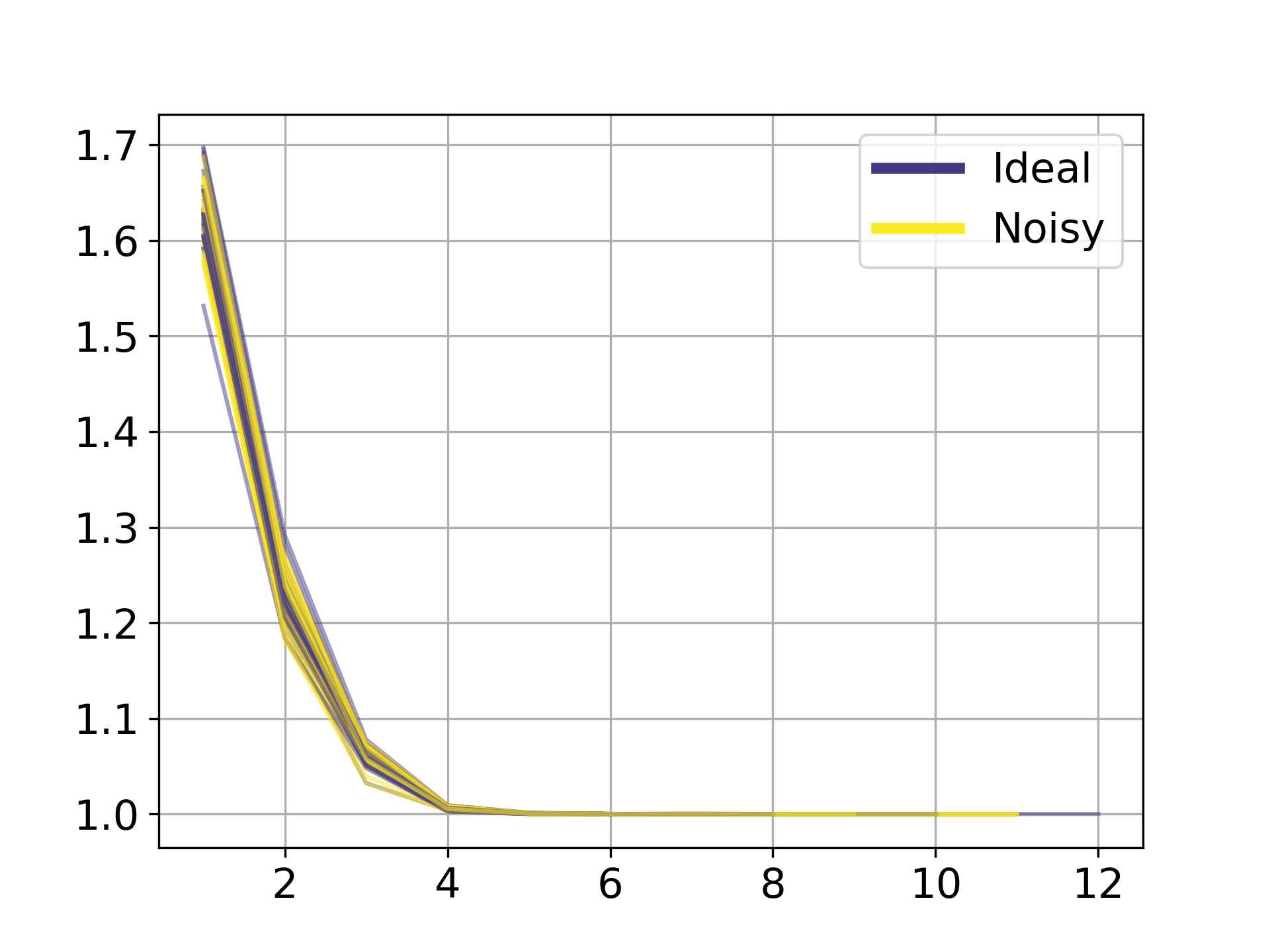}
\caption{\revision{IOF/ROL}}
\label{fig:noise_trend_sub2}
\end{subfigure}
\begin{subfigure}{.19\textwidth}
\centering
\includegraphics[width=1\linewidth]{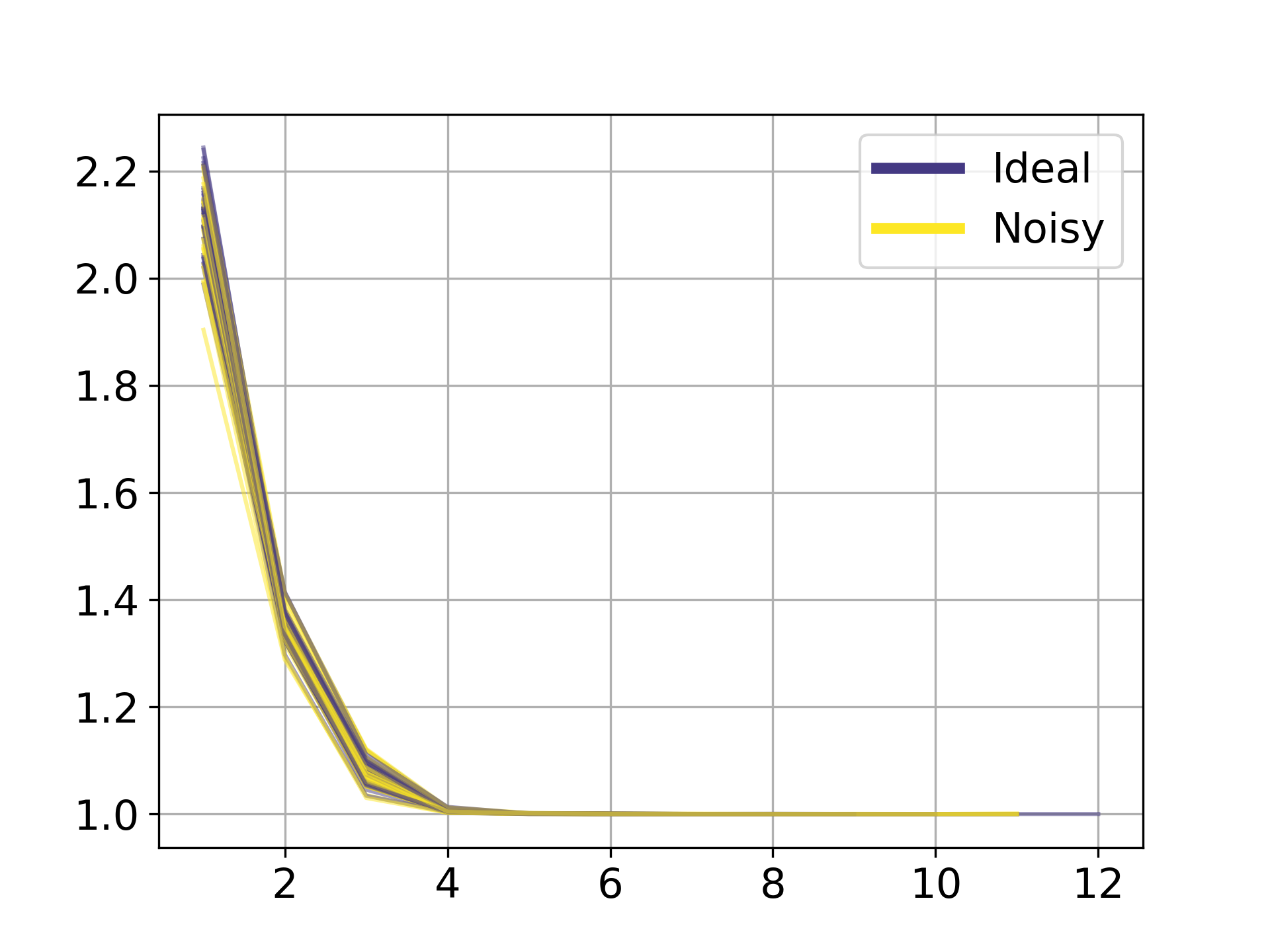}
\caption{\revision{ELEVATOR$_{o2}$}}
\label{fig:noise_trend_sub3}
\end{subfigure}
\begin{subfigure}{.19\textwidth}
\centering
\includegraphics[width=1\linewidth]{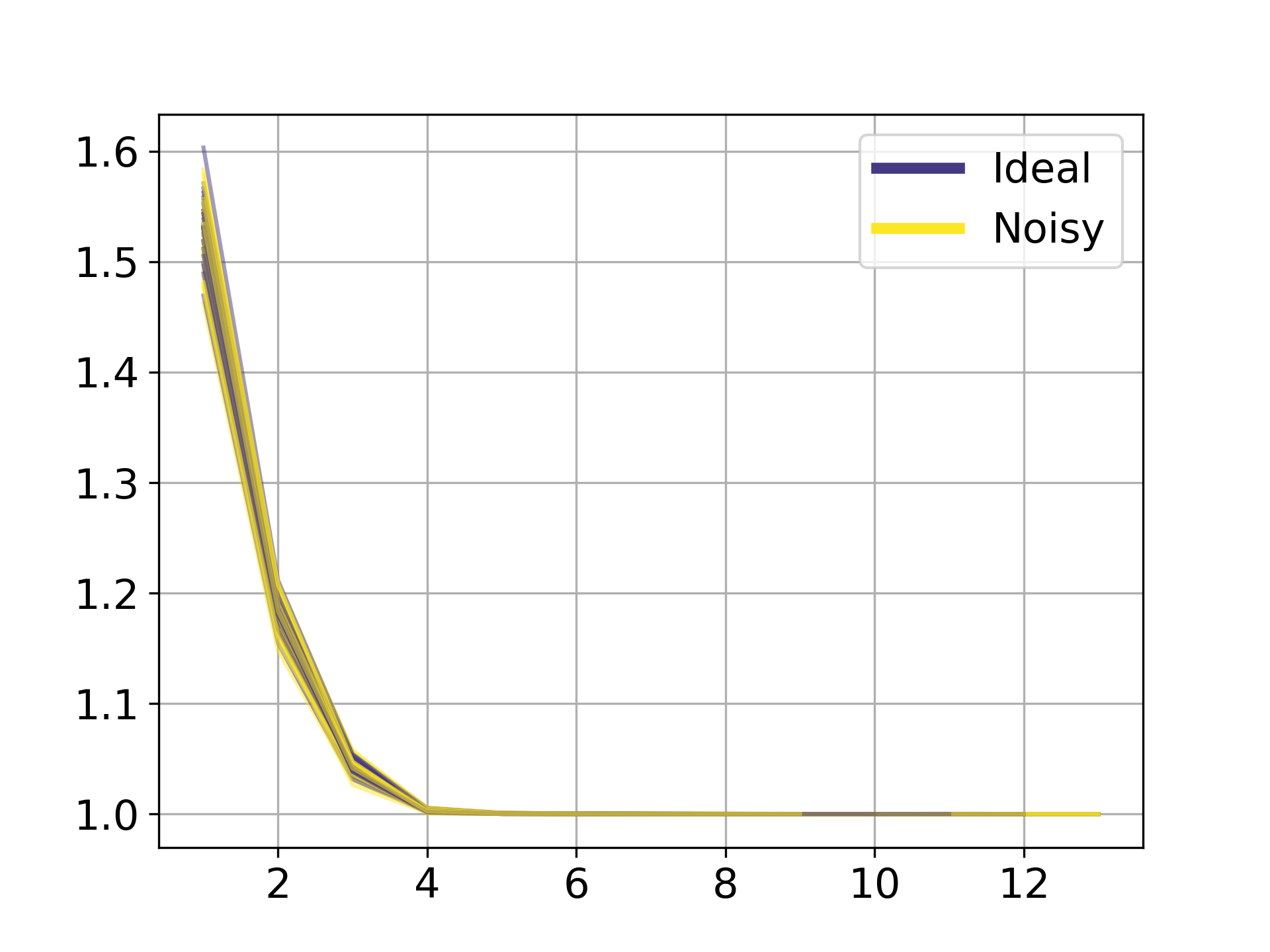}
\caption{\revision{ELEVATOR$_{o3}$}}
\label{fig:noise_trend_sub4}
\end{subfigure}
\begin{subfigure}{.19\textwidth}
\centering
\includegraphics[width=1\linewidth]{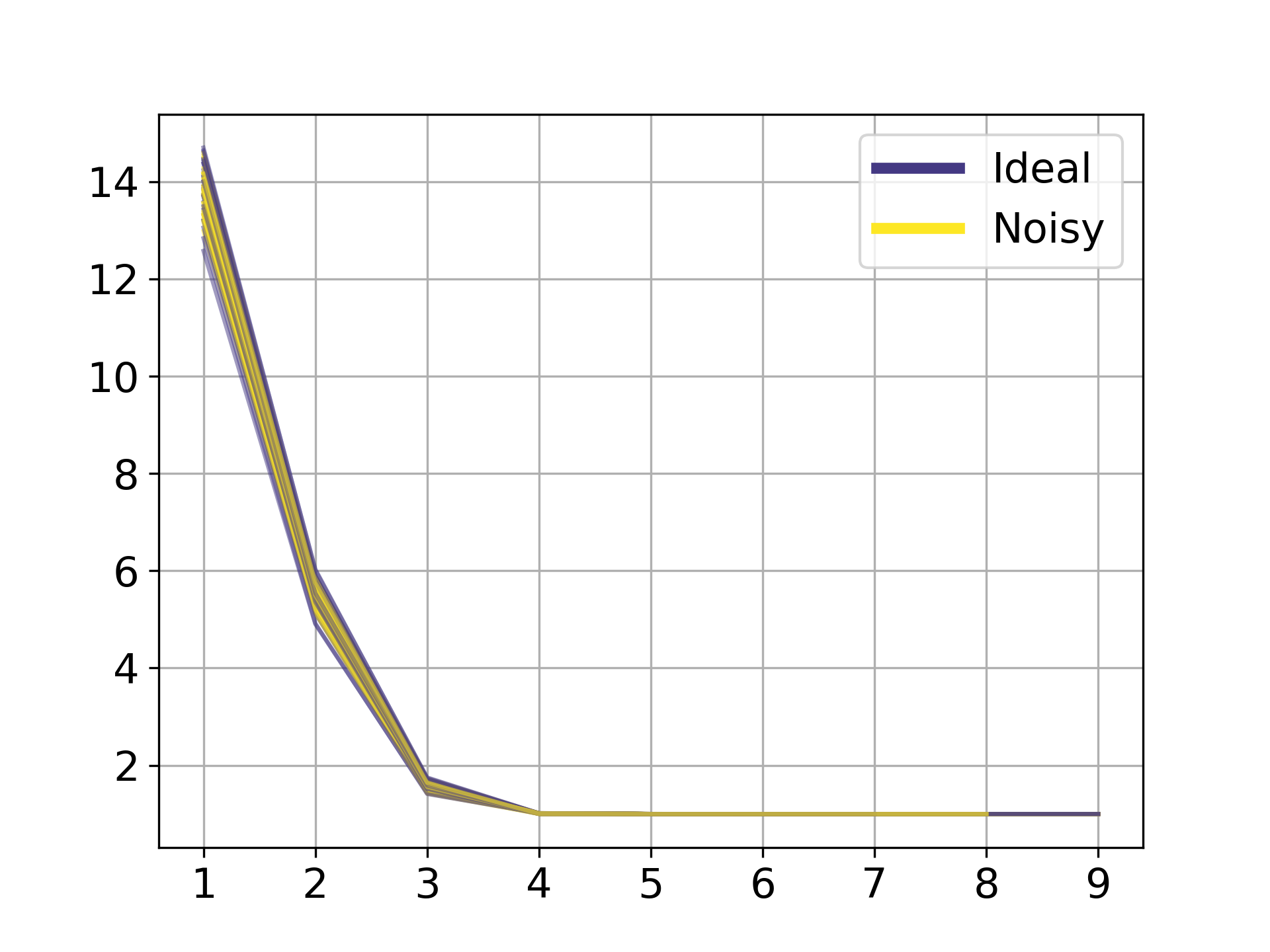}
\caption{\revision{GSDTSR}}
\label{fig:noise_trend_sub5}
\end{subfigure}
\caption{\revision{RQ4 -- Trends of approximation ratios (\appro) along the iterations of \ourApproach on both ideal and noisy simulators}}
\label{fig:noise_trend}
\end{figure*}
The purple line shows the trends of the \appro values achieved by \ourApproach on the ideal simulator, while the yellow lines represent the trends on the noisy simulator. Our observation is that the two trends overlap a lot in all five case studies, indicating that the decline rates \appro values achieved by the two simulators are quite similar, as they all decrease dramatically in the first three iterations and stabilize to near 1.0 in the next few iterations. We employ the Mann-Whitney U test to compare \numEval achieved by \ourApproach on the two simulators. It shows that all p-values obtained in all case studies are larger than 0.05, demonstrating no significant difference between \numEval values achieved by the two simulators. Furthermore, when comparing the solutions achieved, \ourApproach implemented on both simulators consistently finds the optimal solution in all case studies except for \textit{Elevator$_{o3}$}. According to Sect.~\ref{subsec:rq4}, \ourApproach with the ideal simulator fails to find the optimal solution in 10 runs, while \ourApproach with the noisy simulator fails in 12 runs. We also conduct the statistical test in terms of \appro achieved by two simulators. Mann-Whitney U test indicates that the p-value is greater than 0.05. All these observations demonstrate that, since QAOA is resilient to noise~\cite{farhi2014quantum}, the effect of noise on \ourApproach is not significant.}

\noindent\revision{\textbf{Performance of \ourApproach on quantum computer}}
\revision{Due to the limitation of quantum resources, we perform the TCM problem on a small case study extracted from dataset \pa. Since we use the best setting of \ourApproach in the paper ($\totalVariable=7$ and $p=1$), we select 46 test cases to ensure one optimized sub-problem covers 15\% (see in Sect.~\ref{subsubsec:parameters}) of the case study. We use the K-means clustering algorithm to select representative test cases from the dataset. We implement our experiment on the ``ibm\_brisbane'' quantum computer in the IBM platform. We first use the ideal simulator to solve TCM and consider the lowest fitness value achieved as the optimal. Next, we run \ourApproach on the quantum computer. We observe that the \appro value keeps decreasing from the initial 2.46 to the optimal value of 1.0 in five iterations, as \ourApproach finds the optimal solution on the 5th iteration, which demonstrates the feasibility of implementing \ourApproach on the real quantum computer in the presence of noise.}

\begin{tcolorbox}[size=title, colframe=green!10, width=1\linewidth, colback=green!10,breakable]
\textbf{Conclusions for RQ4:} 
\revision{Since QAOA is a noise-resilient quantum algorithm, the effect of noise on \ourApproach is not significant when implemented on a noisy quantum simulator. In addition, it is also feasible to implement \ourApproach on the quantum computer in the presence of noise.} 
\end{tcolorbox}


\subsection{Concluding Remarks}\label{subsec:concluding}
\revision{In this section, we discuss the effectiveness (Sect.~\ref{sec:effectiveness}) and efficiency (Sect.~\ref{sec:effincency}) of \ourApproach over datasets of different sizes, and how it could be possibly adapted to other software engineering problems (Sect.~\ref{sec:applicability}). 
}

\subsubsection{\revision{Effectiveness of \ourApproach}}\label{sec:effectiveness}
This paper investigates using QAOA, a quantum-classical algorithm, for solving test case minimization and selection problems for classical software systems. Our empirical evaluation based on four industrial datasets of five case studies reveals that even at low-cost configurations (i.e., the shortest depth and smallest sub-problem size), \ourApproach obtained equivalent or better approximate optimal solutions compared to GA. \revision{Our results showed that \ourApproach achieved compatible performance compared to the classical approach GA, and in two cases (i.e., IOF/ROL and ELEVATOR$_{o2}$) outperformed them in terms of effectiveness. The sizes of datasets used in \ourApproach range from 90 to 5555 data points, demonstrating its capability to handle both small and large datasets effectively. This versatility in performance across diverse dataset sizes indicates the robustness of our approach, suggesting that it maintains effectiveness regardless of the dataset scale, and our approach is likely to generalize well beyond the specific datasets used in this study. Nonetheless, additional experiments are needed in the future to explore the stability of \ourApproach.}

\subsubsection{\revision{Efficiency of \ourApproach}}\label{sec:effincency}
Recall, from RQ1, that the execution time (i.e., \extime) of \ourApproach remains manageable for large-scale industrial problems.
However, for GA, as shown in Table~\ref{tab:extime_ga}, the average optimization time of GA with the optimal \pop reaching the lowest fitness values ranges from 0.4 to 1081.5 seconds. 
\begin{table}[!tb]
\caption{Execution time of GA}
\resizebox{\columnwidth}{!}{
\begin{tabular}{c|ccccc}
\toprule
\textbf{Case} & Paint Control & IOF/ROL & ELEVATOR$_{o2}$ & ELEVATOR$_{o3}$ & GSDTSR \\
\midrule
\extime (s) & \revision{0.4} & \revision{146.7} & \revision{232.1} & \revision{279.1} & \revision{1081.5}\\
\bottomrule
\end{tabular}
}
\label{tab:extime_ga}
\end{table}
The results show that the execution time of GA is of a similar order of magnitude to that of \ourApproach, with both being in the order of hundreds of seconds (except for Paint Control, for which GA is more efficient). For the largest case study GSDTSR, the average \extime taken by GA is 1081.5 seconds, higher than that of \ourApproach (\revision{540.0} seconds). In general, the time performance of GA and \ourApproach are comparable. However, with a real quantum computer, the time performance of \ourApproach is expected to improve significantly. 

\revision{
The QAOA is theoretically considered to be a promising algorithm in terms of both effectiveness and efficiency~\cite{blekos2024review}. Generally, the performance of QAOA improves with the increment of depth $p$ (which results in higher cost). However, with the smallest depth $p=1$, QAOA is proved to reach a guaranteed minimum approximation ratio and cannot be efficiently simulated by any classical computer~\cite{farhi2014quantum,farhi2016quantum}. Here we conduct the theoretical analysis of the efficiency of \ourApproach. \ourApproach contains an initialization (i.e., \textbf{Step 1} in Sect.~\ref{subsec:overview}) and an iterative decomposition strategy IGDec (i.e., \textbf{Step 2} and \textbf{Step 3} in Sect.~\ref{subsec:overview}). Let $\totalTestCase$ represent the total number of test cases available.
\textbf{Step 1} randomly assigns values to all decision variables related to test cases to get the initial fitness value, whose time complexity is $\mathcal{O}(\totalTestCase)$.
In \textbf{Step 2}, we flip each test case's variable value and calculate its impact on the objective function while keeping others constant. The time complexity is $\mathcal{O}(\totalTestCase)$. Then, we implement a quick sort for all test cases to order the impact, whose time complexity is $\mathcal{O}(\totalTestCase\log\totalTestCase)$. In \textbf{Step 3}, assume that the sub-problem size is predefined as \totalVariable. Since the number of test cases that we optimize with QAOA is $\mathit{num} = 0.15\totalTestCase$ (see Section.~\ref{sec:expDesign}), the number of QAOA executions we implement in one iteration is $0.15\totalTestCase/\totalVariable$. The execution time of QAOA depends on sub-problem size \totalVariable and the depth $p$ selected. We represent the time complexity of QAOA as $\mathcal{O}(f(\totalVariable,p))$. Thus, the time complexity of \textbf{Step 3} is $\mathcal{O}(\totalTestCase f(\totalVariable, p)/\totalVariable)$.
Assuming the number of iterations of IGDec as \textit{r}, the total time complexity of \ourApproach is $\mathcal{O}(\totalTestCase)+\mathcal{O}(r\totalTestCase)+\mathcal{O}(r\totalTestCase\log\totalTestCase)+\mathcal{O}(r\totalTestCase f(\totalVariable, p)/\totalVariable)$.}

\revision{In our experiment, for the iterative steps, \textbf{Step 2} and \textbf{Step 3}, since we empirically limit the maximum number of iterations for \method to 30 (see Section.~\ref{sec:expDesign}), we treat the number of iterations \textit{r} as a constant. We also observe that both the sub-problem size \totalVariable and the number of layers $p$ are independent of $\totalTestCase$, as we achieve comparable results of GA in all case studies with setting a fixed configuration $\totalVariable = 7$ and $p=1$ respectively. As a result, $\mathcal{O}(f(\totalVariable, p))$ remains constant and the time complexity \textbf{Step 3} is $\mathcal{O}(\totalTestCase)$. Thus, the overall time complexity of \method with our configuration is $\mathcal{O}(\totalTestCase\log\totalTestCase)$, which is considered to be efficient.}

\subsubsection{\revision{Possible Further Applications of \ourApproach}}\label{sec:applicability}
These results encourage further investigation for other TCO problems (such as test case prioritization) and even other software engineering optimization problems, e.g., optimally allocating requirements for reviewing, finding optimal configurations for highly configurable systems, and searching for optimal software refactoring methods~\cite{ramirez2019survey}. The way of creating the Ising formulations for such problems will be very similar to what we present in this paper for TCO problems. \revision{For example, for resource allocation problems, assigning a resource to a specific task can be represented as a decision variable, and the suitability of that resource to the task can be considered the corresponding attribute value. We optimize the assignment of various resources to all tasks to maximize the total suitability~\cite{otero2009systematic}.} In the future, we foresee the need to recommend optimal Ising formulations for each kind of software engineering problem and identify suitable settings for QAOA and the impact-guided decomposition strategy of \ourApproach, via empirical studies.

\subsection{Threats to Validity} \label{subsec:threats}
Generalizing results to datasets other than the ones used in this paper is a common threat to the validity of our experiments. \revision{However, the sizes of the five industrial TCO problems range from 90 to 5555, allowing for an evaluation of approaches across diverse problem scales. It exhibited the potential of applicability of \ourApproach on other datasets with similar characteristics as ours. In addition, the Ising formulation proposed in our work is general to be applied to other similar search-based software engineering problems (mentioned in Sect.~\ref{subsec:concluding}), which indicates a potential direction for further exploration.} 
Still, additional experiments are warranted with more diverse datasets of varied sizes while also considering the limited number of qubits we could simulate on classical and real quantum computers currently. \revision{As quantum computing technology evolves, future research could explore the scalability and applicability of our methods to larger and more varied datasets.}

For the baseline algorithms, we used the default parameter settings of GA and RS from jMetalPy. Default parameters have been demonstrated to give good results~\cite{ParameterTunning}. We empirically chose the optimal population size of GA for each case study. For \ourApproach, we had to configure some parameters, i.e., number of layers $p$, sub-problem size $N$, and \fractionNum. We chose $p$ based on our experiment results and \fractionNum based on \textit{qbsolv} from D-Wave~\cite{dwavePartitioning}. For \subproblemSize, we chose it to be 7, 8, 10, 12, 14, and 16, considering the time budget for the quantum simulator execution and the existing qubit ranges of small-scale near-term quantum computers.

To deal with randomness, we repeated each experiment 30 times. Following, we also employed statistical tests, i.e., the Kruskal–Wallis test to study differences among more than two samples, whereas the Mann Whitney U Test and \Atwelve for comparing two samples based on existing guide~\cite{arcuri2011practical}. 

Another threat is the choice of the termination criterion, which is predefined and can incorporate human biases. To alleviate this problem, we choose the termination criterion empirically for each algorithm to achieve its best performance. We \revision{set the maximum iteration number to 30, and within these 30 runs, we} stop \ourApproach, when there is no improvement in fitness for three consecutive generations. \revision{Note that the number of iterations needed to find the optimal solution varies depending on the specific optimization of the problem and is not necessarily achieved within 30 iterations. However, since all our experiments concluded within 30 iterations, we have set this as the maximum limit}. For GA, we observed the evolution trend in each case study and ran 400,000 evaluations in total to ensure that GA converges to the optimal solution. For RS, we generate solutions randomly until the number of solutions reaches \numEval of \ourApproach with $p=1$ and $\subproblemSize=7$.

\section{Related Work}\label{sec:related}
\noindent\textbf{\revision{Classical Test Case Optimization:}} 
Given the scarcity of software testing resources, it is common to apply test optimization techniques to optimize testing resources while at the same time not compromising testing effectiveness. \revision{Some surveys summarize research on test case optimization techniques, e.g., test case minimization (TCM), test case selection (TCS), and test case prioritization (TCP)~\cite{Survey1, xiao2023systematic}. Among those works, search-based approaches are commonly applied, such as genetic algorithm~\cite{wang2015cost}, simulated annealing~\cite{tcs1}, and NSGA-II~\cite{ARRIETA2019137}. Given the popularity of machine learning (ML), several approaches for test case optimization have been published~\cite{survey2} (e.g., using supervised learning~\cite{bertolino2020learning, chen2018optimizing, mahdieh2020incorporating}, unsupervised learning~\cite{khalid2019weight,kandil2017cluster}, reinforcement learning~\cite{bertolino2020learning, bagherzadeh2021reinforcement}, and natural language processing~\cite{kandil2017cluster, medhat2020framework}). However, as the scale of software increases, large classical computational resources become necessary to explore the significant part of the search space of possible test case optimization solutions, posing computational challenges for classical algorithms. Thus, we explore integrating quantum computational resources to tackle this challenge. In particular, we investigate using a quantum optimization algorithm, i.e., QAOA, to solve TCO problems on quantum computers.}

\noindent\textbf{\revision{Quantum Algorithms for Solving SE problems:}}
\revision{With the continuous QC development, its applications in various domains are increasing, such as chemistry, finance, and cryptography~\cite{hassija2020forthcoming}. In recent years, several quantum algorithms have been proposed to solve software engineering (SE) problems. Miranskyy \etal~\cite{miranskyy2022quantum} examined the potential of applying eight groups of quantum algorithms (e.g., quantum machine learning, combinatorial optimization) to SE problems throughout the software development lifecycle(e.g., code quality estimation and release problems). Miranskyy \etal~\cite{miranskyy2022using} also used Grover's algorithm and quantum counting algorithm to speed up the process of dynamic testing on classical programs. Jhaveri \etal~\cite{jhaveri2023cloning} proposed an approach to explore quantum annealing to code clone detection problems. Especially, there has been work on TCO problems. For example, Hussei~\cite{hussein2021quantum} proposed a quantum algorithm to solve TCM problems. This approach contains two stages: first, preparing an incomplete superposition of the search space, and second, finding the solution using an updated version of Arima's algorithm~\cite{arima2009proposal}. Results show that the algorithm can find a solution with high probability in $O(\sqrt{2^n})$ where $n$ represents the number of test cases. This paper's TCM problem is formulated as an SAT problem and is demonstrated theoretically using a 4-test case example. In contrast, \ourApproach is designed to solve large-scale, realistic TCO problems. Bajaj \etal~\cite{bajaj2022test}proposed a quantum-behaved particle swarm optimization combined with GA to solve the TCO problem. This algorithm is inspired by the quantum behavior of particles and QC is not involved. Wang \etal~\cite{wang2023test} propose a hybrid approach to solve TCM problems with random sampling and quantum annealing, which uses different problem formulations and different technology.}

\noindent\textbf{\revision{Applications of QAOA:}}
The computations performed by current quantum computers are noisy, thereby limiting their successful applications. To this end, hybrid algorithms are being advocated to maximize the use of scarce quantum resources and combine them with classical computations in near-term quantum computing applications. QAOA is one such hybrid algorithm that focuses on solving combinatorial optimization problems~\cite{farhi2014quantum}. \revision{It owns the potential of computational advantages\cite{farhi2016quantum,guerreschi2019qaoa, blekos2024review}. It has been proved to be applicable to finding approximated solutions to many well-known optimization problems, such as MaxCut problems, maximum independent set problems, Knapsack problems, job shop scheduling problems, and portfolio optimization problems~\cite{blekos2024review}. Some works have also proved the speedup potential of QAOA. For instance, Zhang \etal~\cite{zhang2022applying} apply QAOA to solve the minimum vertex cover problems in polynomial time with a high probability, which realizes exponential acceleration. Jiang \etal~\cite{jiang2017near} obtain a quadratic speedup over classical algorithms, which is the same as Grover search when solving an unstructured search problem with QAOA. They demonstrate the quantum advantage even with multiple layers (i.e., $p$) in QAOA. For solving quantum linear system problems, An \etal~\cite{an2022quantum} find that QAOA with an optimal control protocol outperforms the best classical baseline algorithm in terms of runtime performance with exponential advantage. It also outperforms several quantum algorithms such as HHL algorithm~\cite{harrow2009quantum}. Real-world applications have also been coming out recently. For instance, Vikstål \etal~\cite{vikstaal2020applying} applies QAOA to an airline scheduling problem, tail-assignment problems, with small-scale real-world data. Results indicate that QAOA can solve the problems with high success probabilities even with low layers (i.e., $p$). In the computer vision domain, Li \etal~\cite{li2020hierarchical} implement QAOA in the partially occluded object detection problem. They achieve execution speedup by choosing a proper classical optimizer, exploiting parameter symmetry, and rescheduling gate operation. Niroula \etal~\cite{niroula2022constrained} construct extractive summarization as an optimization problem and formulate the objective as a quadratic binary optimization problem (QUBO). They implement a QAOA variant, XY-QAOA, on a trapped-ion quantum computer and emphasize the necessity of encoding the constraints into the quantum circuit. In the financial industry, Baker \etal ~\cite{baker2022wasserstein} benchmark the performance of QAOA with various quantum backends on financial assets portfolio optimization problems. Date \etal~\cite{date2021qubo} use QAOA for accelerating the process of training machine learning models. They formulate the training problems of three machine learning models as QUBOs, and they show that both the time and space complexities of their formulations are better than those of their classical counterparts. Despite so much investigation into QAOA, there is a lack of investigation into using QAOA to solve software testing problems. Our work combines the IGDec strategy and QAOA to target a large-scale realistic TCO problem. Such investigation is novel.}

\section{Conclusion and Future Work}\label{sec:conclusion}
Quantum Computing (QC) holds great potential to solve many important and challenging problems in the long term. In the near future, their applications to solve optimization problems seem realistic. To this end, Quantum Approximate Optimization Algorithms (QAOAs) are emerging as promising algorithms with several applications demonstrated in areas such as finance and scheduling. Along this line of investigation, we studied QAOAs in the context of test optimization for software systems. We contributed by providing a generic formulation of test optimization to solve it with QAOAs. Moreover, we integrated an existing test impact order strategy with QAOAs such that they can solve larger problems that cannot be solved with today's limited number of available qubits. We performed an empirical study with five case studies and four industrial datasets. \revision{First, we empirically evaluated \ourApproach on an ideal simulator.} The results demonstrated the benefit of using \ourApproach for solving test optimization problems, which managed to reach similar or even better effectiveness compared to a classical genetic algorithm. \revision{Second, our experiments on a noisy simulator demonstrated that the impact of noise on \ourApproach is insignificant, thus making it a viable solution for real quantum computers.} In the future, we want to investigate solving other software engineering optimization problems with QAOAs.

\ifCLASSOPTIONcompsoc
\section*{Acknowledgments}
\else
\section*{Acknowledgment}
\fi

This work is supported by the Qu-Test project (\#299827) funded by the Research Council of Norway. X. Wang is also supported by Simula's internal strategic project on quantum software engineering. S. Ali also acknowledges the support from the \textit{Quantum Hub initiative} (OsloMet). P. Arcaini is supported by Engineerable AI Techniques for Practical Applications of High-Quality Machine Learning-based Systems Project (Grant Number JPMJMI20B8), JST-Mirai; and by ERATO HASUO Metamathematics for Systems Design Project (No. JPMJER1603), JST, Funding Reference number: 10.13039/501100009024 ERATO.

\ifCLASSOPTIONcaptionsoff
\newpage
\fi



%


\bibliographystyle{IEEEtran}
\bibliography{biblio}

%

\end{document}